# Electrically tunable topological transport of moiré polaritons

Hongyi Yu[1,2†], Wang Yao[2,3*]

[1] Guangdong Provincial Key Laboratory of Quantum Metrology and Sensing & School of Physics and Astronomy, Sun Yat-Sen University (Zhuhai Campus), Zhuhai 519082, China

[2] Department of Physics, The University of Hong Kong, Hong Kong, China

[3] HKU-UCAS Joint Institute of Theoretical and Computational Physics at Hong Kong, China

[†] yuhy33@mail.sysu.edu.cn

[*] wangyao@hku.hk

**Abstract:** Moiré interlayer exciton in transition metal dichalcogenide heterobilayer features a permanent electric dipole that enables the electrostatic control of its flow, and optical dipole that is spatially varying in the moiré landscape. We show the hybridization of moiré interlayer exciton with photons in a planar 2D cavity leads to two types of moiré polaritons that exhibit distinct forms of topological transport phenomena including the spin/valley Hall and polarization Hall effects, which feature remarkable electrical tunability through the control of exciton-cavity detuning by the interlayer bias.

## 1. Introduction

Exciton-polaritons are half-light, half-matter quantum particles formed through the hybridization of photon with exciton in semiconductors, which are of key scientific and technological interest [1, 2]. Atomically thin transition metal dichalcogenides (TMDs) have become a new area to explore polariton physics with excitons in the truly 2D limit [3, 4], exploiting the strong light coupling and the exotic physics of exciton's valley pseudospin [5-18].

Topological properties of polariton arise from its internal quantum degrees of freedom. The dependence of photon's polarization vectors on momentum introduces a non-Abelian gauge structure that leads to a polarization-dependent anomalous velocity upon passing through medium of inhomogeneous refractive index, i.e. the Hall effect of light [19, 20]. With optical selection rule setting correspondence between spin or valley pseudospin of exciton to the polarization of photon in their hybridization [21, 22], polaritons can inherit this gauge structure, and the resultant polariton spin/valley Hall effects are discovered in III-V semiconductors and monolayer TMDs in microcavities [14, 23-26]. With the absence of electric coupling, temperature or density gradient can be exploited to drive the flow of the neutral polariton in these systems [27]. For optoelectronic applications, however, it is desirable to have electrical control on the spin/valley transport of these light-matter hybrid particles.

Interlayer exciton (IX) in TMDs heterobilayers features a permanent electric dipole that enables the desired electrostatic control to direct its flow [28-31]. This electric dipole further enables the gate tunability of IX detuning from the cavity mode for engineering the polariton dispersions, as well as stronger exciton-exciton interactions [32]. Remarkably, the formation of moiré pattern due to the lattice mismatch and twisting of the heterobilayers leads to a superlattice landscape that endows IX an extra pseudospin, characterizing its trapping at different locales with distinct optical dipoles in a moiré supercell [33, 34]. Various signatures of these moiré IX are recently observed in experiments [31, 35-40]. Numerical calculations and experimental observations both show that the IX optical dipole is about one order of magnitude smaller than that of the monolayer ones [33, 40-43]. From the measured ~50 meV Rabi splitting values of monolayer exciton-polaritons [5-17], we estimate that the coupling strength between the IX and the microcavity photon can reach several meV, comparable to the IX linewidth in hBN encapsulated heterobilayers [31]. With the possibility to reach strong coupling of IX with microcavity photon, the remarkable properties of moiré IX point to a desirable new polariton system.

Here we analyze the fine structures of polaritons formed by the hybridization of moiré IXs in a TMD heterobilayer with photons in a planar 2D cavity. We show that in a moiré superlattice potential there are three IX eigenmodes per valley that feature left-handed in-plane, right-handed in-plane, and out-of-plane optical dipoles respectively, which can be selectively coupled to cavity TE and TM modes with the conjugating mode spatial profiles and polarizations. These lead to the possibility to engineer two types of moiré polaritons that inherit gauge structures in different ways from the photon constituents. These moiré polaritons exhibit distinct forms of topological transport phenomena including the spin/valley Hall and polarization Hall effects, which feature remarkable electrical tunability through the control of IX-cavity detuning by the interlayer bias.

## 2. IX in a moiré superlattice and its coupling to light

The optical response of monolayer TMDs is dominated by the Wannier-type intralayer excitons with the momentum space distribution localized near the hexagonal Brillouin zone corners, namely the $\pm \mathbf{K}$ valleys. The two valleys are related by a time reversal and have spin-valley locking due to the large spin-orbit splitting (~30 meV for the conduction band and 100-500 meV for the valence band) [44]. Van der Waals heterobilayers of TMDs can be formed from two different monolayers. Its band structure has a type-II alignment, i.e., conduction and valence band edges are located in different layers thus the low energy exciton has an interlayer configuration. A small lattice mismatch $\delta \equiv |a - a'|/a \ll 1$ and/or twist angle $\delta\theta \ll 1$ between the two layers can lead to the formation of a large-scale moiré superlattice pattern, with a periodicity $\approx a/\sqrt{\delta^2 + \delta\theta^2}$. Here $a$ and $a'$ are the lattice constants of the monolayers. The atomic registry of a local region much smaller than the moiré periodicity can be well described by a commensurate TMD bilayer with certain interlayer translation, which varies smoothly across the moiré supercell. This then introduces a local-to-local variation to the heterobilayer band structure [33, 43].

We consider here a near R-type MoSe$_2$/WSe$_2$ heterobilayer moiré pattern as example, which is formed through vertically stacking a MoSe$_2$ monolayer and a WSe$_2$ monolayer with a twist angle ~ 0°. The low energy IX in the moiré superlattice can be described by $\widehat{H}_X = \widehat{H}_{X,0} + V_{\text{moiré}}(\mathbf{r}_e, \mathbf{r}_h)$. Here $\widehat{H}_{X,0}$ describes the binding of an electron-hole pair by the Coulomb attraction, and $V_{\text{moiré}}(\mathbf{r}_e, \mathbf{r}_h)$ is the moiré induced local-to-local variation of the conduction and valence bands with $\mathbf{r}_{e/h}$ being the electron/hole spatial coordinate. Since the band structure modulation has a spatial scale much larger than the exciton Bohr radius, below we shall first consider $\widehat{H}_{X,0}$, and then add the effect of $V_{\text{moiré}}(\mathbf{r}_e, \mathbf{r}_h)$ as a perturbation.

Under the effective mass approximation,

$$\widehat{H}_{X,0} = \frac{\hbar^2}{2m_e}\left(-i\frac{\partial}{\partial \mathbf{r}_e} - \tau'\mathbf{K}'\right)^2 + \frac{\hbar^2}{2m_h}\left(-i\frac{\partial}{\partial \mathbf{r}_h} + \tau\mathbf{K}\right)^2 + V_C(\mathbf{r}_e - \mathbf{r}_h).$$

Here $-i\frac{\partial}{\partial \mathbf{r}_e} - \tau'\mathbf{K}'$ is the kinematic momentum of an electron in the $\tau'\mathbf{K}'$ valley of MoSe$_2$, and $-i\frac{\partial}{\partial \mathbf{r}_h} + \tau\mathbf{K}$ is that of a hole in the $\tau\mathbf{K}$ valley of WSe$_2$. $m_e$ ($m_h$) is the effective mass near $\tau'\mathbf{K}'$ ($\tau\mathbf{K}$) and $\tau', \tau = \pm 1$. The Coulomb interaction $V_C$ in the heterobilayer results in strongly-bound IXs with exceptionally large binding energies of several hundred meV [45-47], and small Bohr radii of $a_B \sim 2$ nm [48].

It is shown that the electron-hole kinematic momentum sum $\mathbf{Q}$ (the eigenvalue of $\left(-i\frac{\partial}{\partial \mathbf{r}_e} - \tau'\mathbf{K}'\right) + \left(-i\frac{\partial}{\partial \mathbf{r}_h} + \tau\mathbf{K}\right)$) is conserved by the Coulomb interaction, the eigenstate $X_{\tau'\tau,\mathbf{Q}}$ of $\widehat{H}_{X,0}$ is then denoted as the kinematic momentum eigenstate, with an eigenenergy $E_X + \frac{\hbar^2 Q^2}{2M_0}$ [34, 41]. Meanwhile, the momentum conservation in the exciton-photon coupling process requires the IX crystal momentum, $\mathbf{Q} + \tau'\mathbf{K}' - \tau\mathbf{K}$ (the eigenvalue of $-i\frac{\partial}{\partial \mathbf{r}_e} - i\frac{\partial}{\partial \mathbf{r}_h}$), to match the photon wave vector which can be taken as zero at the energy concerned.

With the discrete translational symmetry in the moiré, a reciprocal lattice vector of the moiré superlattice can also be supplied to assist the momentum conservation. Writing the primitive reciprocal lattice vectors of MoSe$_2$ (WSe$_2$) as $\mathbf{G}'_{1,2}$ ($\mathbf{G}_{1,2}$) with $|\mathbf{G}'_1 - \mathbf{G}_1| = |\mathbf{G}'_2 - \mathbf{G}_2| \ll G_1$, the primitive reciprocal lattice vectors of the moiré superlattice are $\mathbf{b}_1 \equiv \mathbf{G}'_1 - \mathbf{G}_1$ and $\mathbf{b}_2 \equiv \mathbf{G}'_2 - \mathbf{G}_2$. The optically bright momentum eigenstates then have $\tau' = \tau$ and $\mathbf{Q}$ centered at $\tau(\boldsymbol{\kappa} - \boldsymbol{\kappa}')$ with $\boldsymbol{\kappa}' \equiv \mathbf{K}' + m\mathbf{G}'_1 + n\mathbf{G}'_2$ and $\boldsymbol{\kappa} \equiv \mathbf{K} + m\mathbf{G}_1 + n\mathbf{G}_2$ ($m, n$ are integers). These $\tau(\boldsymbol{\kappa} - \boldsymbol{\kappa}')$, i.e. the light cones, are nonzero values because of the displacement between electron $\tau\mathbf{K}'$ and hole $\tau\mathbf{K}$ valleys (c.f. Fig. 1a). Although they form an ordered array separated by the reciprocal lattice vectors of the moiré superlattice, but only the main light cones at $\tau\mathbf{K}_m$, $\tau\hat{C}_3\mathbf{K}_m$ and $\tau\hat{C}_3^2\mathbf{K}_m$ ($\mathbf{K}_m \equiv \mathbf{K} - \mathbf{K}'$, $\hat{C}_3$ is the $2\pi/3$-rotation operator) need to be kept considering their strongest light coupling (Fig. 1b). The others with larger kinetic energies $\frac{\hbar^2}{2M_0}|\boldsymbol{\kappa} - \boldsymbol{\kappa}'|^2 \geq 4\frac{\hbar^2 K_m^2}{2M_0}$ correspond to Umklapp recombination with orders of magnitude weaker optical dipoles [41].

We focus here on polaritons formed with the spin-singlet moiré IX, while the spin-triplet has a similar structure at a different energy [34]. With the spin-valley locking in 2D TMDs [44], the IX with $\tau' = \tau = +$ ($\tau' = \tau = -$) has spin up (down), and will be denoted hereafter as $X_{\uparrow,\mathbf{Q}} \equiv X_{++,\mathbf{Q}}$ ($X_{\downarrow,\mathbf{Q}} \equiv X_{--,\mathbf{Q}}$). The spin-up (-down) bright IXs at the three main light cones are then $\alpha_{\uparrow,\mathbf{k}} \equiv X_{\uparrow,\mathbf{K}_m+\mathbf{k}}$, $\beta_{\uparrow,\mathbf{k}} \equiv X_{\uparrow,\hat{C}_3\mathbf{K}_m+\mathbf{k}}$ and $\gamma_{\uparrow,\mathbf{k}} \equiv X_{\uparrow,\hat{C}_3^2\mathbf{K}_m+\mathbf{k}}$ ($\alpha_{\downarrow,\mathbf{k}} \equiv X_{\downarrow,-\mathbf{K}_m+\mathbf{k}}$, $\beta_{\downarrow,\mathbf{k}} \equiv X_{\downarrow,-\hat{C}_3\mathbf{K}_m+\mathbf{k}}$ and $\gamma_{\downarrow,\mathbf{k}} \equiv X_{\downarrow,-\hat{C}_3^2\mathbf{K}_m+\mathbf{k}}$) with $k < E_X/\hbar c \sim 10^{-3}$ nm$^{-1}$ that is the size of the light cone, see Fig. 1b. Note that an IX with $\mathbf{Q} \neq \mathbf{0}$ are not $\hat{C}_3$-symmetric since it has a center-of-mass velocity $\propto \mathbf{Q}$, in general its photon emission should be elliptically polarized. Meanwhile the three main light cones are related by the $\hat{C}_3$-rotation, the optical dipoles of $\alpha_\uparrow$, $\beta_\uparrow$ and $\gamma_\uparrow$ then have the forms [41]

$$\mathbf{D}_{\alpha\uparrow} = D_+\mathbf{e}_+^* + D_-\mathbf{e}_-^* + D_z\mathbf{e}_z,$$
$$\mathbf{D}_{\beta\uparrow} = D_+\mathbf{e}_+^* + e^{-i\frac{2\pi}{3}}D_-\mathbf{e}_-^* + e^{-i\frac{4\pi}{3}}D_z\mathbf{e}_z, \qquad (1)$$
$$\mathbf{D}_{\gamma\uparrow} = D_+\mathbf{e}_+^* + e^{-i\frac{4\pi}{3}}D_-\mathbf{e}_-^* + e^{-i\frac{2\pi}{3}}D_z\mathbf{e}_z.$$

Here $\mathbf{e}_\pm \equiv (\mathbf{e}_x \pm i\mathbf{e}_y)/\sqrt{2}$ with $\mathbf{e}_x$, $\mathbf{e}_y$ and $\mathbf{e}_z$ the three Cartesian unit vectors. $\mathbf{e}_\pm^*$ and $\mathbf{e}_z$ components of the optical dipole couple to the in-plane left-/right-handed circular ($\sigma_\pm$) and vertical linear (z) polarized photon, respectively. Those with spin-down are the time reversal of Eq. (1).

The IX also feel a moiré induced local-to-local variation of the band structure, i.e. $V_{\text{moiré}}(\mathbf{r}_e, \mathbf{r}_h)$. Since the spatial scale of this variation (determined by the moiré period, usually in the range between several to several tens nm) is much larger than the IX Bohr radius, the effect of $V_{\text{moiré}}(\mathbf{r}_e, \mathbf{r}_h)$ can be approximated by a periodic potential $V(\mathbf{R})$ applied on the IX center-of-mass degree of freedom, with $\mathbf{R} \equiv \frac{m_e\mathbf{r}_e + m_h\mathbf{r}_h}{M_0}$ the center-of-mass coordinate and $M_0 \equiv m_e + m_h$ the exciton mass. $V(\mathbf{R})$ corresponds to the local gap between the conduction and valence bands, and has been obtained using first-principles calculations and analytical fits [33] (Fig. 1c). Such a periodic potential can be expanded into the Fourier series. Keeping only the leading Fourier components, $V(\mathbf{R}) = V_0(e^{i\mathbf{b}_1\cdot\mathbf{R}} + e^{i\mathbf{b}_2\cdot\mathbf{R}} + e^{i\mathbf{b}_3\cdot\mathbf{R}}) + V_0^*(e^{-i\mathbf{b}_1\cdot\mathbf{R}} + e^{-i\mathbf{b}_2\cdot\mathbf{R}} + e^{-i\mathbf{b}_3\cdot\mathbf{R}})$, with $V_0$ the complex Fourier amplitude, $\mathbf{b}_{1,2}$ the primitive reciprocal lattice vectors of the moiré, and $\mathbf{b}_3 = -\mathbf{b}_1 - \mathbf{b}_2$. $V(\mathbf{R})$ then couples a main light cone to its six nearest-neighbor light cones with a strength $V_0$ (Fig. 1b). The Hamiltonian of moiré IXs at the main light cones becomes

$$\hat{H}_X = \left(E_X + \frac{\hbar^2 K_m^2}{2M_0}\right) \sum_{\sigma=\uparrow,\downarrow} (|\alpha_{\sigma,\mathbf{k}}\rangle\langle\alpha_{\sigma,\mathbf{k}}| + |\beta_{\sigma,\mathbf{k}}\rangle\langle\beta_{\sigma,\mathbf{k}}| + |\gamma_{\sigma,\mathbf{k}}\rangle\langle\gamma_{\sigma,\mathbf{k}}|)$$
$$+ V_0 \sum_{\sigma=\uparrow,\downarrow} (|\beta_{\sigma,\mathbf{k}}\rangle\langle\alpha_{\sigma,\mathbf{k}}| + |\gamma_{\sigma,\mathbf{k}}\rangle\langle\beta_{\sigma,\mathbf{k}}| + |\alpha_{\sigma,\mathbf{k}}\rangle\langle\gamma_{\sigma,\mathbf{k}}|) + h.c.,$$

where $E_X$ can be dramatically tuned by a vertical electric field. For R-type MoSe$_2$/WSe$_2$ heterobilayers, $E_X \sim 1.4$ eV at zero field, and *ab initio* calculations give $V_0 = (-2.0 - 8.9i)$ meV [33]. For twist angles $\gtrsim 3°$, the energy difference between the Umklapp and main light cones is $\geq 3\frac{\hbar^2 K_m^2}{2M_0} \geq$ **50 meV** (Fig. 1d), much larger than $|V_0|$, so the Umklapp light cones can be well neglected.

$\widehat{H}_X$ has three branches of eigenstates (Fig. 2a):

$$|A_{\sigma,\mathbf{k}}\rangle = \frac{1}{\sqrt{3}}(|\alpha_{\sigma,\mathbf{k}}\rangle + |\beta_{\sigma,\mathbf{k}}\rangle + |\gamma_{\sigma,\mathbf{k}}\rangle),$$
$$|B_{\sigma,\mathbf{k}}\rangle = \frac{1}{\sqrt{3}}\left(|\alpha_{\sigma,\mathbf{k}}\rangle + e^{\sigma\frac{2\pi}{3}i}|\beta_{\sigma,\mathbf{k}}\rangle + e^{\sigma\frac{4\pi}{3}i}|\gamma_{\sigma,\mathbf{k}}\rangle\right), \qquad (2)$$
$$|C_{\sigma,\mathbf{k}}\rangle = \frac{1}{\sqrt{3}}\left(|\alpha_{\sigma,\mathbf{k}}\rangle + e^{\sigma\frac{4\pi}{3}i}|\beta_{\sigma,\mathbf{k}}\rangle + e^{\sigma\frac{2\pi}{3}i}|\gamma_{\sigma,\mathbf{k}}\rangle\right),$$

with well separated energies: $E_A = E_X + \frac{\hbar^2 K_m^2}{2M_0} + V_0 + V_0^*$, $E_B = E_X + \frac{\hbar^2 K_m^2}{2M_0} + V_0 e^{-\frac{2\pi}{3}i} + V_0^* e^{\frac{2\pi}{3}i}$, $E_C = E_X + \frac{\hbar^2 K_m^2}{2M_0} + V_0 e^{\frac{2\pi}{3}i} + V_0^* e^{-\frac{2\pi}{3}i}$. $\sigma = +1/-1$ correspond to ↑/↓. For $V_0 = (-2.0 - 8.9i)$ meV, $E_A - E_B \approx 9.4$ meV and $E_C - E_A \approx 21.4$ meV, see Fig. 2a. It is straightforward to get their optical dipoles from Eq. (1):

$$\mathbf{D}_{A\uparrow} = \sqrt{3}D_+\mathbf{e}_+^*, \qquad \mathbf{D}_{B\uparrow} = \sqrt{3}D_-\mathbf{e}_-^*, \qquad \mathbf{D}_{C\uparrow} = \sqrt{3}D_z\mathbf{e}_z. \qquad (3)$$

$A_\uparrow$ and $B_\uparrow$ couple to photons with in-plane circular polarizations $\sigma_+$ and $\sigma_-$, respectively. Whereas $C_\uparrow$ couples to a photon with z polarization (Fig. 2a). In these eigenmodes in the moiré potential, the IX probability distribution is a periodic function of $\mathbf{R}$ maximized at $A$, $B$ and $C$ locales in Fig. 1c respectively.

We emphasize that the three distinct IX eigenmodes $A_\sigma$, $B_\sigma$ and $C_\sigma$ originate from the twisting and/or lattice-mismatch induced displacement between $\mathbf{K}'$ and $\mathbf{K}$. Such a displacement results in the three different main light cones $\tau \mathbf{K}_m$, $\tau \hat{C}_3 \mathbf{K}_m$ and $\tau \hat{C}_3^2 \mathbf{K}_m$ ($\mathbf{K}_m \equiv \mathbf{K} - \mathbf{K}'$) for each valley, as well as the emergence of a moiré superlattice. The distinct polarization selection rules of the three eigenmodes come from the local $\hat{C}_3$ symmetry at $A$, $B$ and $C$ in the moiré supercell [33]. On the other hand, in a lattice-matched heterobilayer with $\mathbf{K}' = \mathbf{K}$ all light cones merge into a single one at $\mathbf{Q} = 0$, thus there is only one bright eigenmode with zero kinetic energy. In contrast to $A_\sigma$, $B_\sigma$ and $C_\sigma$ eigenmodes of the moiré superlattice, the optical dipole in a lattice-matched heterobilayer depends sensitively on the interlayer registry, and generally contains all $\mathbf{e}_+^*$, $\mathbf{e}_-^*$ and $\mathbf{e}_z$ components.

## 3. Moiré IX-polaritons and their topological properties

Here we analyze the coupling between the moiré IX and the cavity photon. Considering the distinct optical dipoles of $A_\sigma$, $B_\sigma$ and $C_\sigma$, we need to analyze the electric field profile of the cavity mode first. A 2D planar cavity hosts two types of modes with the mode profiles (Fig. 2b, also see the Supplemental Materials for details)

$$\mathbf{u}_{TE,\mathbf{k}} = \sqrt{\frac{2}{L}}\sin(k_z z)(-\sin\theta_\mathbf{k}\,\mathbf{e}_x + \cos\theta_\mathbf{k}\,\mathbf{e}_y), \qquad (4)$$

$$\mathbf{u}_{TM,\mathbf{k}} = \sqrt{\frac{2}{L}}\left(\frac{k_z \sin(k_z z)}{(k_z^2+k^2)^{1/2}}\cos\theta_\mathbf{k}\, \mathbf{e}_x + \frac{k_z \sin(k_z z)}{(k_z^2+k^2)^{1/2}}\sin\theta_\mathbf{k}\, \mathbf{e}_y - i\frac{k\cos(k_z z)}{(k_z^2+k^2)^{1/2}}\mathbf{e}_z\right).$$

$\mathbf{k} \equiv k(\cos\theta_\mathbf{k}, \sin\theta_\mathbf{k})$ is the in-plane wave vector. TE mode couples to $A_\sigma$ and $B_\sigma$ IX eigenmodes of the in-plane optical dipoles. TM mode can couple to both the $A_\sigma$, $B_\sigma$ IX eigenmodes and the $C_\sigma$ eigenmode with the out-of-plane dipole. Keeping up to the linear term of $\mathbf{k}$, it is convenient to use the basis $|\pm_\mathbf{k}\rangle = \frac{|TM_\mathbf{k}\rangle \pm i|TE_\mathbf{k}\rangle}{\sqrt{2}}e^{\pm i\theta_\mathbf{k}}$, where the moiré IX-cavity coupling is

$$\widehat{H}_{X-C} = \sin(k_z z)\left(g_A|A_{\uparrow,\mathbf{k}}\rangle + g_B|B_{\downarrow,\mathbf{k}}\rangle\right)\langle +_\mathbf{k}| + \sin(k_z z)\left(g_A|A_{\downarrow,\mathbf{k}}\rangle + g_B|B_{\uparrow,\mathbf{k}}\rangle\right)\langle -_\mathbf{k}|$$
$$+ \cos(k_z z)\hbar k v_C\left(|C_{\uparrow,\mathbf{k}}\rangle + |C_{\downarrow,\mathbf{k}}\rangle\right)\frac{e^{i\theta}\langle +_\mathbf{k}| + e^{-i\theta}\langle -_\mathbf{k}|}{\sqrt{2}} + \text{h.c.}. \quad (5)$$

$g_A$, $g_B$ and $v_C$ are real numbers under a proper gauge choice and proportional to the IX optical dipole, estimated to be about one order of magnitude smaller than that of the monolayer ones [33, 40-43]. From the measured ~50 meV Rabi splitting values of monolayer exciton-polaritons [5-17], we estimate that $g_{A,B}$ can reach several meV, while $v_C$ can be ~ 0.5% of the light speed.

The total Hamiltonian for the coupled moiré IX and cavity is schematically shown in Fig. 2c, featuring remarkable tunability. First the coupling strength $g_{A,B}\sin(k_z z)$ and $\hbar k v_C \cos(k_z z)$ can be tuned by adjusting the heterobilayer vertical position. Second, the detuning between the IXs and the cavity photon can be controlled by an interlayer bias. Below we focus on the two cases where the cavity photon is hybridizing with $A_\sigma$ and $B_\sigma$ IX modes, and with $C_\sigma$ mode, respectively.

When $\cos(k_z z) = 0$, $C_\sigma$ mode of IX is decoupled from the cavity. The Hamiltonian of such AB-polaritons is

$$\widehat{H}_{\text{AB-Pol}} = E_A|A_{\uparrow,\mathbf{k}}\rangle\langle A_{\uparrow,\mathbf{k}}| + E_B|B_{\downarrow,\mathbf{k}}\rangle\langle B_{\downarrow,\mathbf{k}}| + \hbar\omega_\mathbf{k}|+_\mathbf{k}\rangle\langle +_\mathbf{k}| + \left(g_A|A_{\uparrow,\mathbf{k}}\rangle + g_B|B_{\downarrow,\mathbf{k}}\rangle\right)\langle +_\mathbf{k}|$$
$$+E_A|A_{\downarrow,\mathbf{k}}\rangle\langle A_{\downarrow,\mathbf{k}}| + E_B|B_{\uparrow,\mathbf{k}}\rangle\langle B_{\uparrow,\mathbf{k}}| + \hbar\omega_\mathbf{k}|-_\mathbf{k}\rangle\langle -_\mathbf{k}| + \left(g_A|A_{\downarrow,\mathbf{k}}\rangle + g_B|B_{\uparrow,\mathbf{k}}\rangle\right)\langle -_\mathbf{k}|, \quad (6)$$

which can be separated into two decoupled subspaces $\{+_\mathbf{k}, A_{\uparrow,\mathbf{k}}, B_{\downarrow,\mathbf{k}}\}$ and $\{-_\mathbf{k}, A_{\downarrow,\mathbf{k}}, B_{\uparrow,\mathbf{k}}\}$ related by a time reversal. Fig. 3c,d show the calculated AB-polariton dispersions with the three well separated branches. Each branch is two-fold degenerate.

When $\sin(k_z z) = 0$, or when $A_\sigma$ and $B_\sigma$ modes are far red detuned from the cavity (Fig. 2c), we can just focus on the coupling of $C_\sigma$ mode with the cavity photon. The Hamiltonian for such C-polaritons is

$$\widehat{H}_{\text{C-Pol}} = E_C\left(|C_{b,\mathbf{k}}\rangle\langle C_{b,\mathbf{k}}| + |C_{d,\mathbf{k}}\rangle\langle C_{d,\mathbf{k}}|\right) + \hbar\omega_\mathbf{k}\left(|TM_\mathbf{k}\rangle\langle TM_\mathbf{k}| + |TE_\mathbf{k}\rangle\langle TE_\mathbf{k}|\right)$$
$$+\sqrt{2}\hbar k v_C\left(|C_{b,\mathbf{k}}\rangle\langle TM_\mathbf{k}| + |TM_\mathbf{k}\rangle\langle C_{b,\mathbf{k}}|\right). \quad (7)$$

Here $|C_{b,\mathbf{k}}\rangle \equiv (|C_{\uparrow,\mathbf{k}}\rangle + |C_{\downarrow,\mathbf{k}}\rangle)/\sqrt{2}$ couples to the cavity TM mode, whereas $|C_{d,\mathbf{k}}\rangle \equiv (|C_{\uparrow,\mathbf{k}}\rangle - |C_{\downarrow,\mathbf{k}}\rangle)/\sqrt{2}$ is dark. The resulted four branches of eigenstates are the cavity TE

photon (*TE*) and dark IX (*D*), as well as the lower-polariton (*LP*) and upper-polariton (*UP*) that are the hybridizations of the cavity TM photon with the bright IX (Fig. 4a).

When the IX is subject to an in-plane external force **F**, e.g. the gradient of an interlayer bias introduced by a split gate (see Fig. 3a), the polariton's momentum is then driven: $\hbar \frac{d}{dt}\mathbf{k}(t) = \mathcal{F}$. Here $\mathcal{F} = \rho_{n,\mathbf{k}}\mathbf{F}$, $\rho_{n,\mathbf{k}}$ being IX's weighting in the *n*th polariton branch. As the polariton inherits both the IX spin and photon polarization, the driving force can introduce spin and/or polarization dependent transport, originating from the gauge structures of the photonic component (Fig. 3b).

For the AB-polariton, we write the two degenerate eigenstates of the *n*th branch as $|n_{\pm,\mathbf{k}}\rangle$ ($n = 1,2,3$, see Fig. 3c,d). In each AB-polariton branch, the dependence of the internal structure of polariton, i.e. polarization vector of the photonic component, on the momentum **k**, leads to an Abelian Berry curvature $\mathbf{\Omega}_{n,\mathbf{k}}^{(+)} = i\left\langle \frac{\partial u_{n,+,\mathbf{k}}}{\partial \mathbf{k}} \middle| \times \middle| \frac{\partial u_{n,+,\mathbf{k}}}{\partial \mathbf{k}} \right\rangle = \mathbf{e}_z(\rho_{n,\mathbf{k}} - 1)\frac{k_z}{(k_z^2+k^2)^{3/2}} + \mathbf{e}_z \frac{\mathbf{k}\cdot\nabla_\mathbf{k}\rho_{n,\mathbf{k}}}{k_z^2+k^2+k_z\sqrt{k_z^2+k^2}} = -\mathbf{\Omega}_{n,\mathbf{k}}^{(-)}$, with $|u_{n,\pm,\mathbf{k}}\rangle$ the periodic part of $|n_{\pm,\mathbf{k}}\rangle$ (see the Supplemental Materials). The polariton can also inherit the Berry curvature from the IX constituent, but its strength is several orders of magnitude smaller than that from the photon.

The driving force $\mathcal{F}$ then introduces an anomalous velocity $\pm\mathbf{\Omega}_{n,\mathbf{k}}^{(+)} \times \mathcal{F}$. Both the IX spin $\hat{S}_z \equiv |A_{\uparrow,\mathbf{k}}\rangle\langle A_{\uparrow,\mathbf{k}}| - |A_{\downarrow,\mathbf{k}}\rangle\langle A_{\downarrow,\mathbf{k}}| + |B_{\uparrow,\mathbf{k}}\rangle\langle B_{\uparrow,\mathbf{k}}| - |B_{\downarrow,\mathbf{k}}\rangle\langle B_{\downarrow,\mathbf{k}}|$, and photon polarization $\hat{\sigma}_z \equiv |+_\mathbf{k}\rangle\langle +_\mathbf{k}| - |-_\mathbf{k}\rangle\langle -_\mathbf{k}|$ have opposite expectation values in the doubly degenerate $|n_{\pm,\mathbf{k}}\rangle$, so the anomalous velocity gives rise to the spin and polarization Hall currents quantified by $\Omega_{n,\mathbf{k}}^{(S)} = \rho_{n,\mathbf{k}}\langle n_{+,\mathbf{k}}|\hat{S}_z|n_{+,\mathbf{k}}\rangle\mathbf{\Omega}_{n,\mathbf{k}}^{(+)} \cdot \mathbf{e}_z$ and $\Omega_{n,\mathbf{k}}^{(P)} = \rho_{n,\mathbf{k}}\langle n_{+,\mathbf{k}}|\hat{\sigma}_z|n_{+,\mathbf{k}}\rangle\mathbf{\Omega}_{n,\mathbf{k}}^{(+)} \cdot \mathbf{e}_z$, respectively. We show the calculated polariton dispersions in Fig. 3c,d and the corresponding $\Omega_{n,\mathbf{k}}^{(S/P)}$ under different IX-cavity detuning in Fig. 3e-h, which indicates that the spin and polarization Hall conductivities for a certain polariton branch can be widely tuned by varying the IX-cavity detuning.

In contrast, the C-polariton has a vanishing Berry curvature and thus zero anomalous velocity in each branch (see the Supplemental Materials). Nevertheless, due to the **k**-dependence of the TE/TM in-plane polarizations, the momentum-space motion driven by the force can coherently mix TE and TM, leading to a **k**-dependent circular polarization $\langle\hat{\sigma}_z\rangle$. Unlike the anomalous Hall velocity of the AB-polariton, the topological nature of C-polariton manifest in this non-equilibrium polarization $\langle\hat{\sigma}_z\rangle$, which, combined with the group velocity $\frac{1}{\hbar}\frac{\partial E_{n,\mathbf{k}}}{\partial \mathbf{k}}$, results in a polarization current. Analysis shows that the finite Berry curvature of the AB-polariton originates from the **k**-dependence of the vertical electric field for the cavity TM mode (see Eq. (4)), whereas $\langle\hat{\sigma}_z\rangle$ of the C-polariton originates from the **k**-dependence of in-plane directions of TE/TM polarizations (see the Supplemental Materials for details).

For the C-polariton *UP/LP* branch, we find the driving force introduces a photon polarization $\langle\hat{\sigma}_z\rangle_{UP/LP,\mathbf{k}} = \frac{2k_z}{\sqrt{k_z^2+k^2}} \frac{(\mathbf{k}\times\mathbf{F})\cdot\mathbf{e}_z}{k^2} \frac{\rho_{LP,\mathbf{k}}\rho_{UP,\mathbf{k}}}{E_{UP/LP,\mathbf{k}}-\hbar\omega_\mathbf{k}}$ up to the linear order in **F** (see the Supplemental Materials for details). Note that both $\langle\hat{\sigma}_z\rangle_{UP,\mathbf{k}}$ and $\langle\hat{\sigma}_z\rangle_{LP,\mathbf{k}}$ are proportional to $\rho_{LP,\mathbf{k}}\rho_{UP,\mathbf{k}}$ which is the product of the IX and photon fractions. This implies that those strongly hybridized polaritons contribute most significantly to polarization current, as the driving force only affects IX component whereas the gauge structure comes from the photonic component. In Fig. 4b-c we show $\langle\hat{\sigma}_z\rangle_{LP,\mathbf{k}}/F$ and $\langle\hat{\sigma}_z\rangle_{UP,\mathbf{k}}/F$ as functions of **k** when **F** is along the *y*-direction. Note that the above result doesn't apply to $k \to \mathbf{0}$ or very large *k* value. For these two limits either $\hbar\omega_\mathbf{k} - E_{LP,\mathbf{k}} \to \mathbf{0}$ or $E_{UP,\mathbf{k}} - \hbar\omega_\mathbf{k} \to \mathbf{0}$, and our perturbative treatment breaks down.

The total polarization current depends on the polariton **k**-space occupation. If the occupation is isotropic, then the current direction is perpendicular to the driving force, which corresponds to a polarization Hall effect. On the other hand, one can selectively excite a polariton branch at given **k** by choosing the frequency and incident direction of the excitation, and the resultant polarization current can be non-perpendicular to **F**.

The above polarization Hall effect for the C-polariton can also be understood from a spin model under the effect of a time-dependent Zeeman field. Introducing the spin-1 Pauli matrices $\hat{\boldsymbol{\sigma}} = (\hat{\sigma}_x, \hat{\sigma}_y, \hat{\sigma}_z)$ with $\hat{\sigma}_x = \frac{1}{\sqrt{2}}(|+_\mathbf{k}\rangle\langle C_{b,\mathbf{k}}| + |C_{b,\mathbf{k}}\rangle\langle -_\mathbf{k}| + \text{h.c.})$, $\hat{\sigma}_y = \frac{1}{\sqrt{2}}(-i|+_\mathbf{k}\rangle\langle C_{b,\mathbf{k}}| - i|C_{b,\mathbf{k}}\rangle\langle -_\mathbf{k}| + \text{h.c.})$ and $\hat{\sigma}_z = |+_\mathbf{k}\rangle\langle +_\mathbf{k}| - |-_\mathbf{k}\rangle\langle -_\mathbf{k}|$, the C-polariton Hamiltonian can be rewritten as

$$\hat{H}_{C-\text{Pol}} = E_C + \mathbf{B}(t)\cdot\hat{\boldsymbol{\sigma}} + (\hbar\omega_\mathbf{k} - E_C)\hat{\sigma}_z^2. \qquad (8)$$

Here $\mathbf{B}(t) = \hbar v_C \mathbf{k}(t)$ is a time-dependent in-plane Zeeman field with $\hbar\frac{d}{dt}\mathbf{k}(t) = \mathcal{F}$. The last term accounts for the spin anisotropy, and we have dropped the irrelevant dark IX. The dynamics of the spin orientation is described by

$$\hbar\frac{d\langle\hat{\boldsymbol{\sigma}}\rangle}{dt} = \mathbf{B}(t)\times\langle\hat{\boldsymbol{\sigma}}\rangle + i(\hbar\omega_\mathbf{k} - E_C)\langle[\hat{\sigma}_z^2,\hat{\boldsymbol{\sigma}}]\rangle + \alpha\hbar\frac{d\langle\hat{\boldsymbol{\sigma}}\rangle}{dt}\times\langle\hat{\boldsymbol{\sigma}}\rangle.$$

$\alpha$ is a small damping parameter. This model is similar to the 2D electron system with Rashba spin-orbit coupling, where the electron feels a **k**-dependent in-plane Rashba field. When the in-plane Zeeman field varies with time, the spin orientation doesn't fully follow the instantaneous direction of the field but acquires a non-equilibrium z-component $\langle\hat{\sigma}_z\rangle$ [49].

Rayleigh scattering of *UP* and *LP* can also cause $\langle\hat{\sigma}_z\rangle$ to develop in different scattering directions, similar to that observed in a GaAs/AlGaAs microcavity with a finite TM-TE splitting [23-25]. Here an effective TM-TE splitting arises from the coupling of the cavity modes with the $C_\sigma$ mode of IX. The $C_\sigma$ IX can only couple with the TM but not TE modes. The hybridization of $C_\sigma$ IX with cavity TM mode results in the C-polariton *UP* and *LP* branches in Fig. 4a, each with a sizable energy splitting from the cavity TE mode (in the order of Rabi splitting, several meV). These can then be viewed as effective TM-TE splitting. The intrinsic TM-TE splitting of the cavity mode plays a negligible role here, because of its small energy scale (~ 0.1 meV [23-25]).

It is expected that the Rayleigh scattering doesn't change the pseudospin configuration (i.e., the amplitudes of $|+_\mathbf{k}\rangle$, $|-_\mathbf{k}\rangle$, $|C_\mathbf{k}\rangle$), but only changes the momentum of polariton [23, 24]. A polariton initially on an eigenstate is then scattered to the superposition of split eigenstates at a different wave vector, whose time evolution then causes the emergence of a finite $\langle \hat{\sigma}_z \rangle$. In Fig. 4d we show the calculated emission circular polarization as functions of scattering angle $\Delta\theta$ for six initial states (assuming 1 ps cavity photon lifetime). Generally, the polarization generated by Rayleigh scattering shows a superposition of two terms with **sin($\Delta\theta$)** and **sin($2\Delta\theta$)** angular patterns, respectively (see the Supplemental Materials). When $E_{UP,\mathbf{k}} - \hbar\omega_\mathbf{k} = \hbar\omega_\mathbf{k} - E_{LP,\mathbf{k}}$ the **sin($2\Delta\theta$)** term vanishes and $\langle \hat{\sigma}_z \rangle \propto$ **sin($\Delta\theta$)** (Fig. 4d).

## 4. Summary and discussion

We have investigated the coupling between the IX in a transition metal dichalcogenide heterobilayer and the cavity photon. The existence of a moiré superlattice pattern endows three bright IX eigenmodes with distinct optical dipoles, whose hybridization with cavity photons could lead to two types of moiré polaritons. The moiré polaritons inherit gauge structures in different ways from the photon constituents, thus exhibit distinct forms of topological transport phenomena including the spin/valley Hall and polarization Hall effects. Combined with the electrical tunability from IX's permanent electric dipole, the Hall conductivities and the current direction of the moiré polaritons feature remarkable electrostatic control through an interlayer bias.

For polaritons subject to a periodic potential, earlier works have also shown the possibility to engineer polariton bands of nontrivial topological number exploiting an applied Zeeman field and a large TM-TE splitting of cavity mode [50, 51], which leads to protected chiral edge channels. In contrast, the topological transports addressed here are the bulk effects naturally endowed by the moiré pattern, where the moiré modulations of exciton energy and optical dipole result in the unique exciton dispersions and pseudospin splitting for hybridizing with the cavity photon. Compared to the small gaps for the exploitation of the chiral polariton edge channel [50, 51], the energy scale for exploring the bulk topological transport is characterized by the O(10) meV splitting between the polariton branches. This, combined with the large excitonic binding energies in TMDs, implies that the effect can be potentially explored for room temperature operation.

The formation of the moiré polariton requires strong coupling between the IX and the cavity photon, i.e., the coupling strength (estimated to be several meV in MoSe$_2$/WSe$_2$) should be larger than the IX linewidth. The IX optical dipole can be further enhanced through reducing the interlayer distance with pressure or use heterobilayers with strong hybridization between inter- and intralayer excitons [38]. For the linewidth, in contrast to the ~ 2 meV value for intralayer excitons in ultraclean monolayer samples [52, 53], the IX linewidth is found to be ~ 5 meV [31] which should be limited by the inhomogeneous broadening from spatial variation. Better heterobilayer sample qualities can help to further reduce the IX linewidth.

We estimate that the moiré polariton discussed here can be explored over a range of twist angle $\delta\theta$ from **~3°** to **~10°** (with a moiré period from 6 nm to 2 nm). The upper limit is based on the consideration of having a well-defined moiré superlattice (for $\delta\theta$ larger than **10°**, the moiré starts to lose the periodicity). The lower limit is from the following considerations. Note that the energy difference $\Delta E$ between the Umklapp and main light cones is $\propto \delta\theta^2$ (see Fig. 1b&d). For $\delta\theta \gtrsim$ **3°**, $\Delta E \geq$ **50** meV, much larger than the moiré potential induced coupling strength $V_0 \approx$ **9** meV between the Umklapp and main light cones. So the three main light cones (with large optical dipoles) are effectively decoupled from the Umklapp ones (with nearly zero optical dipoles). The main light cones hybridize to form the three $A_\sigma$, $B_\sigma$ and $C_\sigma$ IX eigenmodes, all with large optical dipoles. For smaller $\delta\theta$, the coupling $V_0$ can effectively mix the main and Umklapp light cones, leading to a large number of split eigenmodes with small optical dipoles (each mode having a small fraction of the main light cones). This will complicate the analysis of polariton and may hinder the realization of a strong coupling regime. In the regime $\delta\theta \gtrsim$ **3°**, the energies of $A_\sigma$, $B_\sigma$ and $C_\sigma$ IX eigenmodes are $E_A = E_X + \frac{\hbar^2 K_m^2}{2M_0} + V_0 + V_0^*$, $E_B = E_X + \frac{\hbar^2 K_m^2}{2M_0} + V_0 e^{-\frac{2\pi}{3}i} + V_0^* e^{\frac{2\pi}{3}i}$ and $E_C = E_X + \frac{\hbar^2 K_m^2}{2M_0} + V_0 e^{\frac{2\pi}{3}i} + V_0^* e^{-\frac{2\pi}{3}i}$. Their energy splitting is determined by $V_0$, insensitive to the twist angle. The change of twist angle $\delta\theta$ can result in an overall energy shift through the common $\frac{\hbar^2 K_m^2}{2M_0}$ term for all three branches, but does not change their splitting.

**Acknowledgments:** This work is mainly supported by the Key-Area Research and Development Program of Guangdong Province (2019B030330001), Research Grants Council of Hong Kong (17312916), Croucher Foundation, and Seed Funding for Strategic Interdisciplinary Research Scheme of HKU.

[1]     Deng H, Haug H, Yamamoto Y. Exciton-polariton bose-einstein condensation. Rev Mod Phys, 2010, 82: 1489
[2]     Carusotto I, Ciuti C. Quantum fluids of light. Rev Mod Phys, 2013, 85: 299
[3]     Mak KF, He K, Lee C, et al. Tightly bound trions in monolayer mos2. Nat Mater, 2013, 12: 207-211
[4]     Ross JS, Wu S, Yu H, et al. Electrical control of neutral and charged excitons in a monolayer semiconductor. Nat Commun, 2013, 4: 1474
[5]     Dufferwiel S, Schwarz S, Withers F, et al. Exciton-polaritons in van der waals heterostructures embedded in tunable microcavities. Nat Commun, 2015, 6: 8579
[6]     Liu X, Galfsky T, Sun Z, et al. Strong light-matter coupling in two-dimensional atomic crystals. Nat Photon, 2015, 9: 30-34
[7]     Liu X, Bao W, Li Q, et al. Control of coherently coupled exciton polaritons in monolayer tungsten disulphide. Phys Rev Lett, 2017, 119: 027403

[8]     Dhara S, Chakraborty C, Goodfellow KM, et al. Anomalous dispersion of microcavity trion-polaritons. Nat Phys, 2017, 14: 130-133
[9]     Flatten LC, He Z, Coles DM, et al. Room-temperature exciton-polaritons with two-dimensional ws2. Sci Rep, 2016, 6: 33134
[10]    Wang S, Li S, Chervy T, et al. Coherent coupling of ws2 monolayers with metallic photonic nanostructures at room temperature. Nano Lett, 2016, 16: 4368-4374
[11]    Chen Y-J, Cain JD, Stanev TK, et al. Valley-polarized exciton-polaritons in a monolayer semiconductor. Nat Photon, 2017, 11: 431-435
[12]    Sun Z, Gu J, Ghazaryan A, et al. Optical control of room-temperature valley polaritons. Nat Photon, 2017, 11: 491-496
[13]    Dufferwiel S, Lyons TP, Solnyshkov DD, et al. Valley-addressable polaritons in atomically thin semiconductors. Nat Photon, 2017, 11: 497-501
[14]    Lundt N, Dusanowski Ł, Sedov E, et al. Optical valley hall effect for highly valley-coherent exciton-polaritons in an atomically thin semiconductor. Nat Nanotechnol, 2019, 14: 770-775
[15]    Dufferwiel S, Lyons TP, Solnyshkov DD, et al. Valley coherent exciton-polaritons in a monolayer semiconductor. Nat Commun, 2018, 9: 4797
[16]    Qiu L, Chakraborty C, Dhara S, et al. Room-temperature valley coherence in a polaritonic system. Nat Commun, 2019, 10: 1513
[17]    Sidler M, Back P, Cotlet O, et al. Fermi polaron-polaritons in charge-tunable atomically thin semiconductors. Nat Phys, 2017, 13: 255-261
[18]    Schneider C, Glazov MM, Korn T, et al. Two-dimensional semiconductors in the regime of strong light-matter coupling. Nat Commun, 2018, 9: 2695
[19]    Onoda M, Murakami S, Nagaosa N. Hall effect of light. Phys Rev Lett, 2004, 93: 083901
[20]    Hosten O, Kwiat P. Observation of the spin hall effect of light via weak measurements. Science, 2008, 319: 787-790
[21]    Xu X, Yao W, Xiao D, et al. Spin and pseudospins in layered transition metal dichalcogenides. Nat Phys, 2014, 10: 343-350
[22]    Yu H, Cui X, Xu X, et al. Valley excitons in two-dimensional semiconductors. Natl Sci Rev, 2015, 2: 57-70
[23]    Leyder C, Romanelli M, Karr JP, et al. Observation of the optical spin hall effect. Nat Phys, 2007, 3: 628-631
[24]    Kavokin A, Malpuech G, Glazov M. Optical spin hall effect. Phys Rev Lett, 2005, 95: 136601
[25]    Maragkou M, Richards CE, Ostatnický T, et al. Optical analogue of the spin hall effect in a photonic cavity. Optics Lett, 2011, 36: 1095-1097
[26]    Gutiérrez-Rubio Á, Chirolli L, Martín-Moreno L, et al. Polariton anomalous hall effect in transition-metal dichalcogenides. Phys Rev Lett, 2018, 121: 137402
[27]    Onga M, Zhang Y, Ideue T, et al. Exciton hall effect in monolayer mos2. Nat Mater, 2017, 16: 1193-1197
[28]    Rivera P, Schaibley JR, Jones AM, et al. Observation of long-lived interlayer excitons in monolayer mose2-wse2 heterostructures. Nat Commun, 2015, 6: 6242
[29]    Unuchek D, Ciarrocchi A, Avsar A, et al. Room-temperature electrical control of exciton flux in a van der waals heterostructure. Nature, 2018, 560: 340-344
[30]    High AA, Novitskaya EE, Butov LV, et al. Control of exciton fluxes in an excitonic integrated circuit. Science, 2008, 321: 229-231
[31]    Ciarrocchi A, Unuchek D, Avsar A, et al. Polarization switching and electrical control of interlayer excitons in two-dimensional van der waals heterostructures. Nat Photon, 2019, 13: 131-136

[32] Rivera P, Seyler KL, Yu H, et al. Valley-polarized exciton dynamics in a 2d semiconductor heterostructure. Science, 2016, 351: 688-691
[33] Yu H, Liu G-B, Tang J, et al. Moiré excitons: From programmable quantum emitter arrays to spin-orbit-coupled artificial lattices. Sci Adv, 2017, 3: e1701696
[34] Yu H, Liu G-B, Yao W. Brightened spin-triplet interlayer excitons and optical selection rules in van der waals heterobilayers. 2D Mater, 2018, 5: 035021
[35] Tran K, Moody G, Wu F, et al. Evidence for moiré excitons in van der waals heterostructures. Nature, 2019, 567: 71-75
[36] Seyler KL, Rivera P, Yu H, et al. Signatures of moiré-trapped valley excitons in mose2/wse2 heterobilayers. Nature, 2019, 567: 66-70
[37] Jin C, Regan EC, Yan A, et al. Observation of moiré excitons in wse2/ws2 heterostructure superlattices. Nature, 2019, 567: 76-80
[38] Alexeev EM, Ruiz-Tijerina DA, Danovich M, et al. Resonantly hybridized excitons in moiré superlattices in van der waals heterostructures. Nature, 2019, 567: 81-86
[39] Förg M, Colombier L, Patel RK, et al. Cavity-control of interlayer excitons in van der waals heterostructures. Nat Commun, 2019, 10: 3697
[40] Jin C, Regan EC, Wang D, et al. Identification of spin, valley and moiré quasi-angular momentum of interlayer excitons. Nat Phys, 2019, 15: 1140-1144
[41] Yu H, Wang Y, Tong Q, et al. Anomalous light cones and valley optical selection rules of interlayer excitons in twisted heterobilayers. Phys Rev Lett, 2015, 115: 187002
[42] Ross JS, Rivera P, Schaibley J, et al. Interlayer exciton optoelectronics in a 2d heterostructure p–n junction. Nano Lett, 2017, 17: 638-643
[43] Wu F, Lovorn T, MacDonald AH. Theory of optical absorption by interlayer excitons in transition metal dichalcogenide heterobilayers. Phys Rev B, 2018, 97: 035306
[44] Xiao D, Liu G-B, Feng W, et al. Coupled spin and valley physics in monolayers of mos2 and other group-vi dichalcogenides. Phys Rev Lett, 2012, 108: 196802
[45] Wilson NR, Nguyen PV, Seyler KL, et al. Determination of band offsets, hybridization, and exciton binding in 2d semiconductor heterostructures. Sci Adv, 2017, 3: e1601832
[46] Merkl P, Mooshammer F, Steinleitner P, et al. Ultrafast transition between exciton phases in van der waals heterostructures. Nat Mater, 2019, 18: 691-696
[47] Chiu M-H, Li M-Y, Zhang W, et al. Spectroscopic signatures for interlayer coupling in mos2-wse2 van der waals stacking. ACS Nano, 2014, 8: 9649-9656
[48] Donck MVd, Peeters FM. Interlayer excitons in transition metal dichalcogenide heterostructures. Phys Rev B, 2018, 98: 115104
[49] Sinova J, Culcer D, Niu Q, et al. Universal intrinsic spin hall effect. Phys Rev Lett, 2004, 92: 126603
[50] Klembt S, Harder TH, Egorov OA, et al. Exciton-polariton topological insulator. Nature, 2018, 562: 552-556
[51] Bardyn C-E, Karzig T, Refael G, et al. Topological polaritons and excitons in garden-variety systems. Phys Rev B, 2015, 91: 161413
[52] Ajayi OA, Ardelean JV, Shepard GD, et al. Approaching the intrinsic photoluminescence linewidth in transition metal dichalcogenide monolayers. 2D Mater, 2017, 4: 031011
[53] Cadiz F, Courtade E, Robert C, et al. Excitonic linewidth approaching the homogeneous limit in mos2-based van der waals heterostructures. Phys Rev X, 2017, 7: 021026

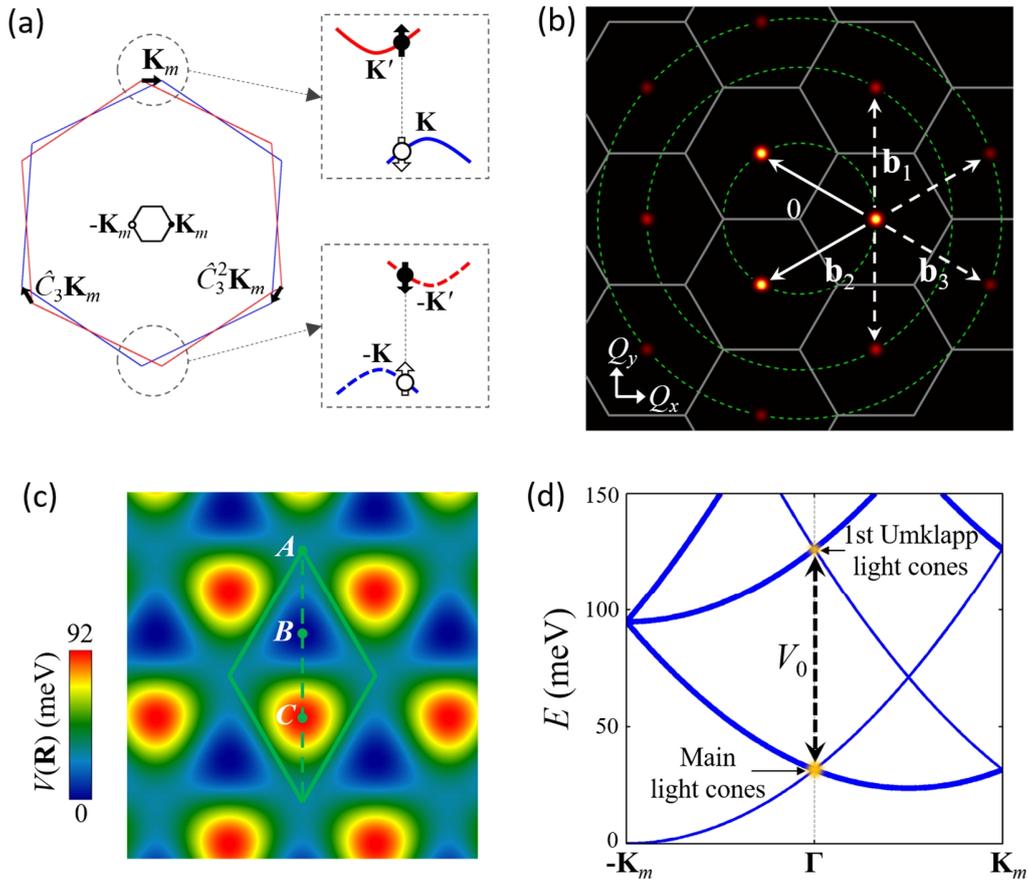

**Fig. 1** (a) The monolayer Brillouin zones of the electron (red) and hole (blue) layers. The black hexagon is the moiré mini Brillouin zone. Insets: electron valley is displaced from hole valley, and therefore an electron-hole pair that can radiatively recombine must have a finite kinematic momentum. (b) Bright spots denote the light cones in the **Q**-space for spin-up IX ($X_\uparrow$), with the three brightest ones for the main light cones. Gray hexagons show the moiré mini BZ, while dashed green circles are the kinetic energy contours. The arrows denote the coupling between light cones introduced by the moiré potential. (c) The moiré potential landscape of a near R-type $MoSe_2/WSe_2$ heterobilayer. *A*, *B* and *C* are the three high-symmetry locales. (d) $X_\uparrow$ dispersion in the absence of moiré potential, folded into the mini Brillouin zone of a 5 nm period moiré pattern. Thick lines are doubly degenerate. All light cones are folded to **Γ** with different energies, which are coupled by the moiré potential.

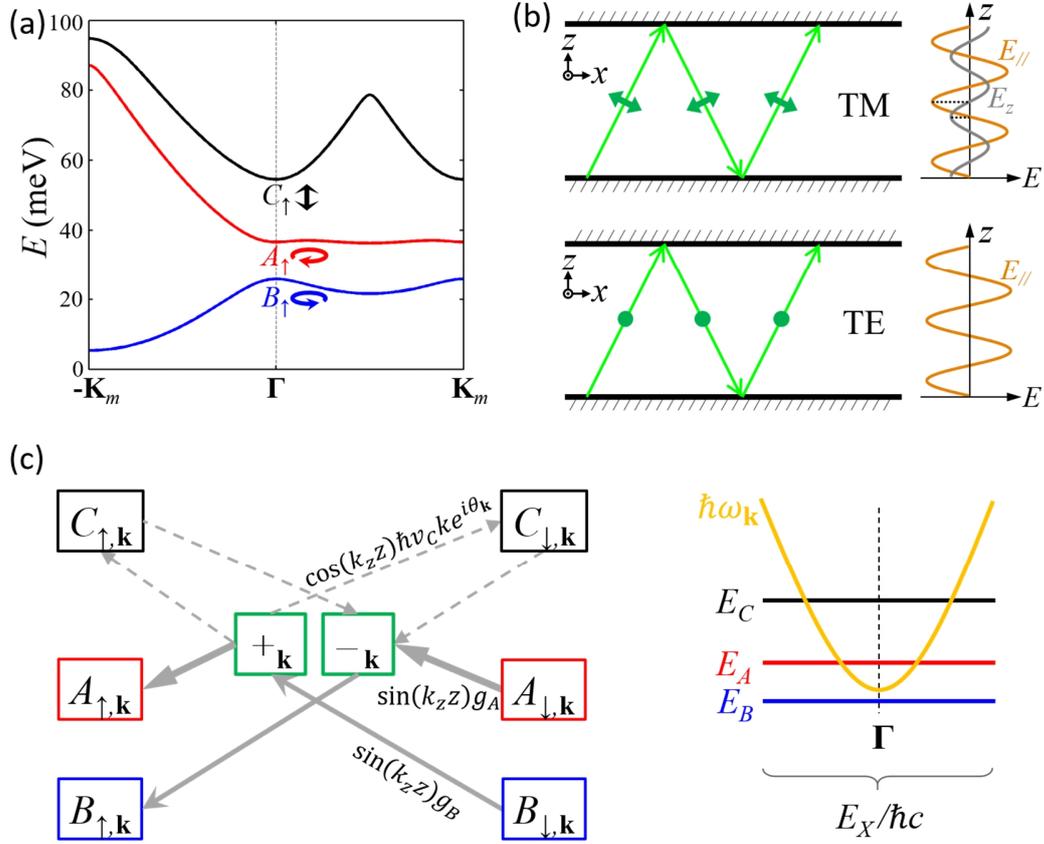

**Fig. 2** (a) The three lowest bands of the spin-up moiré IX with a 5 nm moiré period. The moiré potential leads to three well-separated eigensmodes $B_\uparrow$, $A_\uparrow$ and $C_\uparrow$ at **Γ**, which have $\sigma_-$, $\sigma_+$ and z polarized optical dipoles, respectively. (b) Schematic of TM and TE modes in a planar cavity with momenta along *x*-direction. The in-plane and out-of-plane electric fields have the standing wave profiles $E_\parallel \propto \sin(k_z z)$ and $E_z \propto \cos(k_z z)$ respectively along the confinement direction. (c) Schematic of the coupling between IXs and cavity photons, where the crossing of the cavity dispersion with three branches of IX dispersions can be electrically tuned.

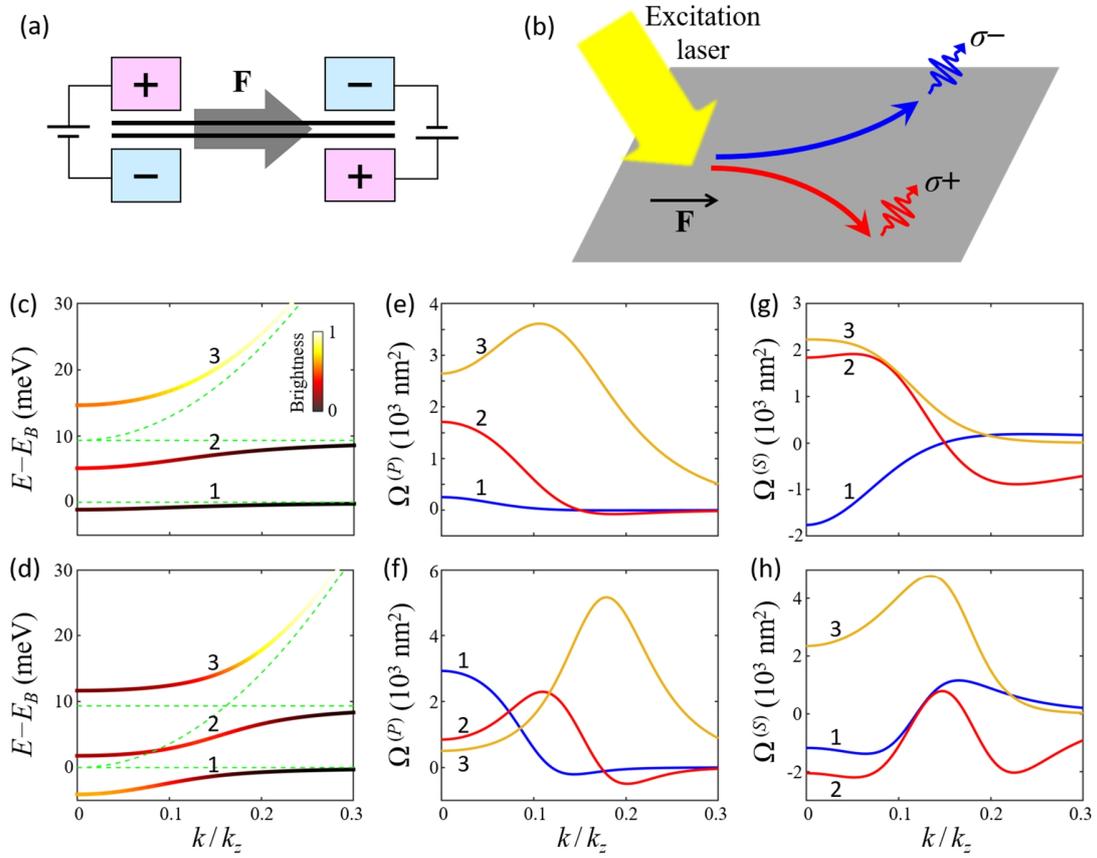

**Fig. 3** (a) The gradient of an interlayer bias introduced by a split gate can apply an in-plane force **F** on the IX. (b) Illustration of the polarization Hall effect driven by **F**. (c,d) The AB-polariton dispersions for two different IX-cavity energy detuning, with coupling strengths $g_A = 5$ meV, $g_B = 3$ meV. The green dashed curves are the cavity mode and $A_\sigma/B_\sigma$ IX dispersions without coupling. The color scale indicates the brightness (photon fraction) of the polariton. (e,f) $\Omega^{(P)}_{n,\mathbf{k}}$, which quantifies the photon polarization Hall response at **k**, for the three polariton branches $n = 1, 2, 3$. (g,h) $\Omega^{(S)}_{n,\mathbf{k}}$, which quantifies the spin Hall response at **k**.

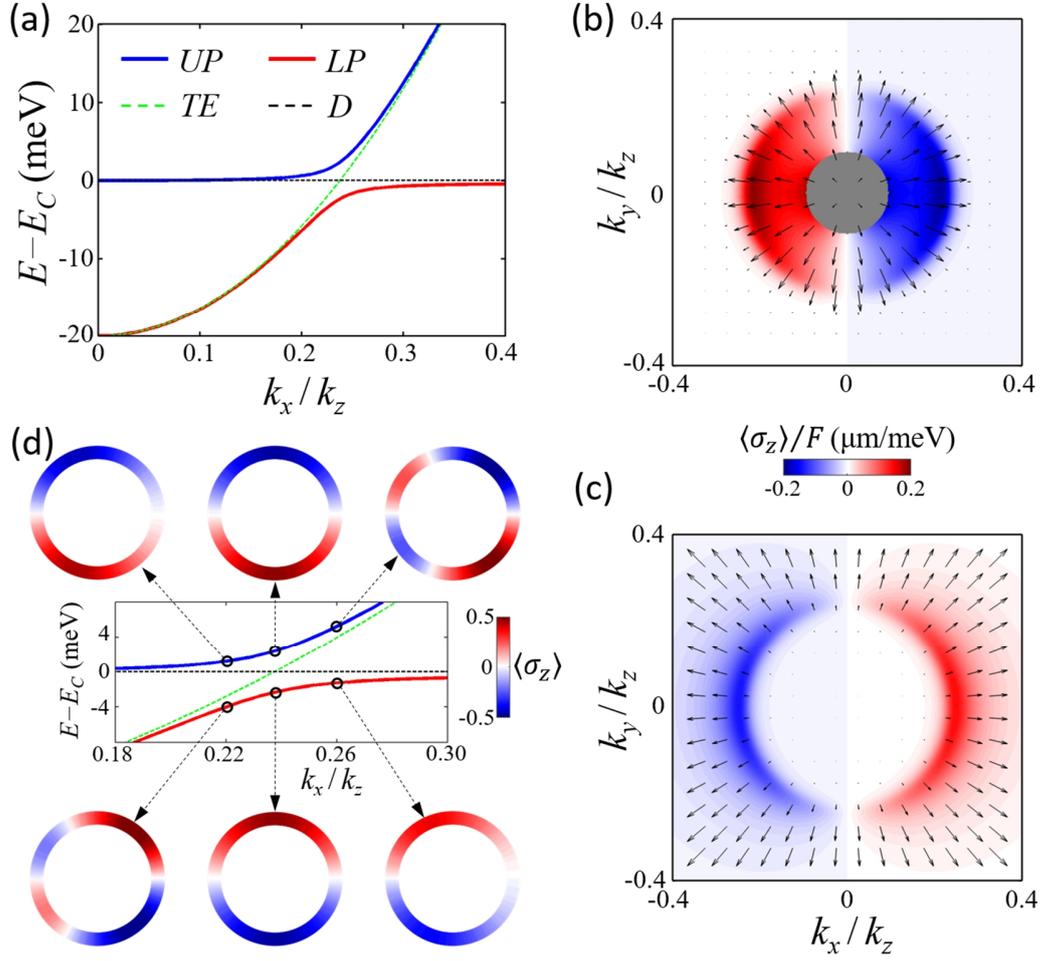

**Fig. 4** (a) The C-polariton dispersions for $v_C = 0.005c$ (c.f. Eq. (7)). (b,c) Photon polarization Hall effect driven by a driving force **F** along $y$ direction. The color map shows **k**-dependent photon polarization $\langle \hat{\sigma}_z \rangle$ induced by **F**, and the arrows indicate the group velocities. (b) is for the *LP* branch, where the results for $k$ close to 0 are not shown since our perturbative treatment breaks down for $\hbar\omega_\mathbf{k} - E_{LP,\mathbf{k}} \to 0$. (c) is for the *UP* branch. (d) Scattering induced polarization currents under excitation at given energy-momentum points (empty dots) in the *LP* and *UP* branches. Cavity photon lifetime is taken as 1 ps.

# Supplemental Materials for
# "Electrically tunable topological transport of moiré polaritons"

**I. Non-Abelian gauge structure of a multi-band system**

We consider a multi-band system in the form of the Bloch functions $\phi_{n,\mathbf{k}}(\mathbf{r}) = e^{i\mathbf{k}\cdot\mathbf{r}}u_{n,\mathbf{k}}(\mathbf{r})$, with $n = 1,2,\ldots,N$ the band index and $\mathbf{k}$ the wave vector. $u_{n,\mathbf{k}}$ is the periodic part which describes the internal structure of the band. These bands are orthogonal but not necessarily to span a complete basis. The momentum-space gauge structure of this multi-band system can be described by the non-Abelian Berry connection, which is in an $N \times N$ matrix form

$$\overleftrightarrow{\mathbf{A}}_{\mathbf{k}} = \begin{pmatrix} \mathbf{A}_{11,\mathbf{k}} & \mathbf{A}_{12,\mathbf{k}} & \cdots & \mathbf{A}_{1N,\mathbf{k}} \\ \mathbf{A}_{21,\mathbf{k}} & \mathbf{A}_{22,\mathbf{k}} & \cdots & \mathbf{A}_{2N,\mathbf{k}} \\ \vdots & \vdots & \ddots & \vdots \\ \mathbf{A}_{N1,\mathbf{k}} & \mathbf{A}_{N2,\mathbf{k}} & \cdots & \mathbf{A}_{NN,\mathbf{k}} \end{pmatrix}, \qquad (S1)$$

with $\mathbf{A}_{mn,\mathbf{k}} = i\left\langle u_{m,\mathbf{k}}\left|\frac{\partial u_{n,\mathbf{k}}}{\partial \mathbf{k}}\right.\right\rangle$.

When these bands are well separated in energy, then under a weak external perturbation the motion of a state can be well restricted to a certain band (i.e., the adiabatic limit). Focusing on this band, it has a gauge-invariant Abelian Berry curvature $\mathbf{\Omega}_{n,\mathbf{k}} = \nabla_{\mathbf{k}} \times \mathbf{A}_{nn,\mathbf{k}} = i\left\langle\frac{\partial u_{n,\mathbf{k}}}{\partial \mathbf{k}}\right| \times \left|\frac{\partial u_{n,\mathbf{k}}}{\partial \mathbf{k}}\right\rangle$, which acts as a magnetic field in $\mathbf{k}$-space. The equation-of-motion for a state with an Abelian Berry curvature has been well studied [1]. It is known that a transverse anomalous velocity $\mathbf{\Omega}_{n,\mathbf{k}} \times \mathbf{F}$ emerges under an external driving force $\mathbf{F}$, which leads to the anomalous Hall effect.

On the other hand if the bands are close in energy or have degeneracies, then we can define a non-Abelian Berry curvature $\overleftrightarrow{\mathbf{\Omega}}_{\mathbf{k}} = \nabla_{\mathbf{k}} \times \overleftrightarrow{\mathbf{A}}_{\mathbf{k}} - i\overleftrightarrow{\mathbf{A}}_{\mathbf{k}} \times \overleftrightarrow{\mathbf{A}}_{\mathbf{k}}$ which is also an $N \times N$ matrix. It's easy to show that, if the considered bands are complete, i.e., $\sum_{n=1}^{N}|u_{n,\mathbf{k}}\rangle\langle u_{n,\mathbf{k}}| = \mathbf{1}$, then the non-Abelian Berry curvature vanishes: $\overleftrightarrow{\mathbf{\Omega}}_{\mathbf{k}} = \mathbf{0}$. Below we briefly analyze the effect of such non-Abelian gauge structure to the dynamics of the state.

Under the perturbation of an external force, the Hamiltonian becomes $\widehat{H}_0 - \mathbf{F}\cdot\hat{\mathbf{r}}$. Here $\widehat{H}_0 = \int d\mathbf{k}\sum_n E_{n,\mathbf{k}}|\phi_{n,\mathbf{k}}\rangle\langle\phi_{n,\mathbf{k}}|$ is the unperturbed Hamiltonian with $E_{n,\mathbf{k}}$ the band energy. The second term contains the position operator $\hat{\mathbf{r}}$, which then requires the knowledge of the $\mathbf{r}$-matrix element $\langle\phi_{m,\mathbf{q}}|\hat{\mathbf{r}}|\phi_{n,\mathbf{k}}\rangle$. From the relation $\frac{\partial\phi_{n,\mathbf{k}}(\mathbf{r})}{\partial\mathbf{k}} = e^{i\mathbf{k}\cdot\mathbf{r}}\frac{\partial u_{n,\mathbf{k}}(\mathbf{r})}{\partial\mathbf{k}} + i\mathbf{r}\phi_{n,\mathbf{k}}(\mathbf{r})$,

$$\langle\phi_{m,\mathbf{q}}|\hat{\mathbf{r}}|\phi_{n,\mathbf{k}}\rangle = i\left\langle u_{m,\mathbf{q}}\left|e^{i(\mathbf{k}-\mathbf{q})\cdot\hat{\mathbf{r}}}\right|\frac{\partial u_{n,\mathbf{k}}}{\partial\mathbf{k}}\right\rangle - i\frac{\partial\langle\phi_{m,\mathbf{q}}|\phi_{n,\mathbf{k}}\rangle}{\partial\mathbf{k}}$$

$$= \delta(\mathbf{k}-\mathbf{q})i\left\langle u_{m,\mathbf{k}}\left|\frac{\partial u_{n,\mathbf{k}}}{\partial\mathbf{k}}\right.\right\rangle - i\delta_{mn}\frac{\partial\delta(\mathbf{k}-\mathbf{q})}{\partial\mathbf{k}}$$

$$= \delta(\mathbf{k}-\mathbf{q})\left(\mathbf{A}_{mn,\mathbf{k}} + i\delta_{mn}\frac{\partial}{\partial\mathbf{k}}\right).$$

A general wavepacket can be constructed as $|\Psi\rangle = \int d\mathbf{k}\sum_n w_{n,\mathbf{k}}|\phi_{n,\mathbf{k}}\rangle$, with $w_{n,\mathbf{k}}$ a set

of state vectors satisfying $\sum_n \int d\mathbf{k} |w_{n,\mathbf{k}}|^2 = 1$ and $\phi_{n,\mathbf{k}}$ the Bloch eigenstates independent on time. Without the external driving force, we expect the wavepacket to be fully on a certain band $n_0$: $|\Psi^{(0)}\rangle = \int d\mathbf{k} \sum_n w_{n,\mathbf{k}}^{(0)} |\phi_{n,\mathbf{k}}\rangle = \int d\mathbf{k} \sqrt{\rho(\mathbf{k})} e^{i\theta_{\mathbf{k}}} |\phi_{n_0,\mathbf{k}}\rangle$ with $w_{n,\mathbf{k}}^{(0)} = \delta_{nn_0}\sqrt{\rho(\mathbf{k})} e^{i\theta_{\mathbf{k}}}$. The density $\rho(\mathbf{k})$ is narrowly distributed near the wavepacket $\mathbf{k}$-space center $\mathbf{k}_c$, such that $\rho(\mathbf{k}) \approx \delta(\mathbf{k} - \mathbf{k}_c)$. The wavepacket real-space center is

$$\langle \Psi^{(0)} | \hat{\mathbf{r}} | \Psi^{(0)} \rangle \approx \mathbf{A}_{n_0 n_0, \mathbf{k}_c} - \frac{\partial \theta_{\mathbf{k}_c}}{\partial \mathbf{k}_c}.$$

A weak driving force $\mathbf{F}$ can lead to finite occupation at other bands with $n \neq n_0$. We focus on the linear response of $\mathbf{F}$ and drop all the $O(F^2)$ terms, then $w_{n_0,\mathbf{k}} \approx w_{n_0,\mathbf{k}}^{(0)}$ and $w_{n,\mathbf{k}} \propto F$ for $n \neq n_0$. We write $w_{n,\mathbf{k}} = \sqrt{\rho(\mathbf{k})} \eta_{n,\mathbf{k}}$, where $\rho(\mathbf{k}) = \sum_n |w_{n,\mathbf{k}}|^2 \approx |w_{n_0,\mathbf{k}}^{(0)}|^2$ corresponds to the wavepacket density distribution. The vector $\vec{\eta}_{\mathbf{k}} \equiv (\eta_{1,\mathbf{k}}, \eta_{2,\mathbf{k}}, \dots, \eta_{N,\mathbf{k}})^T$ can be viewed as the wavepacket pseudospin.

The time evolution of $|\Psi\rangle = \int d\mathbf{k} \sum_n w_{n,\mathbf{k}} |\phi_{n,\mathbf{k}}\rangle$ is governed by the Schrödinger equation $i \frac{\partial |\Psi\rangle}{\partial t} = (\hat{H}_0 - \mathbf{F} \cdot \hat{\mathbf{r}}) |\Psi\rangle$, which then results in

$$i \frac{\partial w_{n,\mathbf{k}}}{\partial t} = E_{n,\mathbf{k}} w_{n,\mathbf{k}} - \mathbf{F} \cdot \left( \sum_m \mathbf{A}_{nm,\mathbf{k}} w_{m,\mathbf{k}} + i \frac{\partial w_{n,\mathbf{k}}}{\partial \mathbf{k}} \right).$$

Keeping up to the linear order of $F$, the above equation can be written as

$$i \frac{\partial w_{n,\mathbf{k}}}{\partial t} = i \eta_{n,\mathbf{k}} \frac{\partial \sqrt{\rho(\mathbf{k})}}{\partial t} + i \sqrt{\rho(\mathbf{k})} \frac{\partial \eta_{n,\mathbf{k}}}{\partial t} \approx E_{n,\mathbf{k}} w_{n,\mathbf{k}} - \mathbf{F} \cdot \left( \sum_m \mathbf{A}_{nm,\mathbf{k}} w_{m,\mathbf{k}}^{(0)} + i \frac{\partial w_{n,\mathbf{k}}^{(0)}}{\partial \mathbf{k}} \right)$$

$$= E_{n,\mathbf{k}} \sqrt{\rho(\mathbf{k})} \eta_{n,\mathbf{k}} - \mathbf{F} \cdot \left( \left( \mathbf{A}_{nn_0,\mathbf{k}} - \delta_{nn_0} \frac{\partial \theta_{\mathbf{k}}}{\partial \mathbf{k}} \right) \sqrt{\rho(\mathbf{k})} e^{i\theta_{\mathbf{k}}} + i \delta_{nn_0} e^{i\theta_{\mathbf{k}}} \frac{\partial \sqrt{\rho(\mathbf{k})}}{\partial \mathbf{k}} \right).$$

Multiplying both sides by $\sqrt{\rho(\mathbf{k})}$ and integrating them over $\mathbf{k}$, we get

$$i \frac{\partial \eta_{n,\mathbf{k}_c}}{\partial t} \approx E_{n,\mathbf{k}_c} \eta_{n,\mathbf{k}_c} - \mathbf{F} \cdot \left( \mathbf{A}_{nn_0,\mathbf{k}_c} - \delta_{nn_0} \frac{\partial \theta_{\mathbf{k}_c}}{\partial \mathbf{k}_c} \right) e^{i\theta_{\mathbf{k}_c}}$$

$$\approx (E_{n,\mathbf{k}_c} - \mathbf{F} \cdot \mathbf{r}_c) \eta_{n,\mathbf{k}_c} - \mathbf{F} \cdot \sum_m (\mathbf{A}_{nm,\mathbf{k}_c} - \delta_{nm} \mathbf{A}_{mm,\mathbf{k}_c}) \eta_{m,\mathbf{k}_c}$$

In the above result we have used the fact that $\rho(\mathbf{k}) \approx \delta(\mathbf{k} - \mathbf{k}_c)$ so $\int d\mathbf{k} \rho(\mathbf{k}) f(\mathbf{k}) \approx f(\mathbf{k}_c)$. Meanwhile $\eta_{m,\mathbf{k}_c} = \delta_{mn_0} e^{i\theta_{\mathbf{k}_c}} + O(F)$, $\mathbf{r}_c = \langle \Psi | \hat{\mathbf{r}} | \Psi \rangle = \mathbf{A}_{n_0 n_0, \mathbf{k}_c} - \frac{\partial \theta_{\mathbf{k}_c}}{\partial \mathbf{k}_c} + O(F)$.

We can see that the pseudospin vector $\vec{\eta}_{\mathbf{k}_c}$ satisfies

$$i \frac{\partial \vec{\eta}_{\mathbf{k}_c}}{\partial t} \approx (\vec{H}_{\mathbf{k}_c} - \mathbf{F} \cdot \mathbf{r}_c - \mathbf{F} \cdot \widetilde{\mathbf{A}}_{\mathbf{k}_c}) \vec{\eta}_{\mathbf{k}_c}. \tag{S2}$$

Here $\overleftrightarrow{H}_{\mathbf{k}_c} \equiv \mathbf{diag}(E_{1,\mathbf{k}_c}, E_{2,\mathbf{k}_c}, \ldots E_{N,\mathbf{k}_c})$ is the diagonal matrix form of the unperturbed Hamiltonian, and $\widetilde{\mathbf{A}}_{\mathbf{k}_c}$ is the off-diagonal parts of the non-Abelian Berry connection $\overleftrightarrow{\mathbf{A}}_{\mathbf{k}_c}$. $\overleftrightarrow{H}_{\mathbf{k}_c} - \mathbf{F} \cdot \mathbf{r}_c - \mathbf{F} \cdot \widetilde{\mathbf{A}}_{\mathbf{k}_c}$ then acts as an effective Hamiltonian which governs the pseudospin dynamics under an external driving force. $\vec{\eta}_{\mathbf{k}_c}$ will eventually fall on an eigenstate of $\overleftrightarrow{H}_{\mathbf{k}_c} - \mathbf{F} \cdot \mathbf{r}_c - \mathbf{F} \cdot \widetilde{\mathbf{A}}_{\mathbf{k}_c}$, as the pseudospin precession gradually damp out. Note that it is the *off-diagonal* parts of the non-Abelian Berry connection, i.e., $\widetilde{\mathbf{A}}_{\mathbf{k}_c}$, that leads to the deviation of the pseudospin to the unperturbed one, even though the non-Abelian Berry curvature can be zero ($\overleftrightarrow{\Omega}_{\mathbf{k}_c} = \mathbf{0}$).

From the solved pseudospin vector $\vec{\eta}_{\mathbf{k}_c}$, the wavepacket center-of-mass motion can be obtained

$$\frac{d\mathbf{k}_c}{dt} = i\langle \Psi | [\widehat{H}_0 - \mathbf{F} \cdot \hat{\mathbf{r}}, \mathbf{k}] | \Psi \rangle = \mathbf{F}, \qquad \text{(S3)}$$

$$\frac{d\mathbf{r}_c}{dt} = i\langle \Psi | [\widehat{H}_0 - \mathbf{F} \cdot \hat{\mathbf{r}}, \hat{\mathbf{r}}] | \Psi \rangle$$

$$\approx \vec{\eta}_{\mathbf{k}_c}^\dagger \left( \frac{\partial \overleftrightarrow{H}_{\mathbf{k}_c}}{\partial \mathbf{k}_c} + i[\overleftrightarrow{H}_{\mathbf{k}_c}, \overleftrightarrow{\mathbf{A}}_{\mathbf{k}_c}] + \overleftrightarrow{\Omega}_{\mathbf{k}_c} \times \mathbf{F} \right) \vec{\eta}_{\mathbf{k}_c}.$$

On the right-hand-side of the last line, the first term is the trivial group velocity ($\approx \frac{\partial E_{n_0,\mathbf{k}_c}}{\partial \mathbf{k}_c}$), and the third term is the anomalous Hall velocity. In the single band ($N = 1$) or fully degenerate case $\overleftrightarrow{H}_{\mathbf{k}_c}$ becomes an identity matrix so the second term vanishes, and the resulted Eq. (S2-S3) is the same as the previously obtained ones [1].

Below we analyze several multi-band systems and give their non-Abelian Berry connections, as well as the corresponding Abelian or non-Abelian Berry curvatures.

*(1) Cavity photon*

For a cavity formed by two perfectly reflecting parallel-plate mirrors, the cavity photon wave vector is quantized in the out-of-plane (*z*) direction, i.e., $k_z = m\pi/L$ with *L* the cavity length and *m* an integer. For each given $k_z$, the electric field **E** inside the cavity can be expanded as

$$\mathbf{E} = \sum_{\mathbf{k}} \left( \mathbf{E}_{TE,\mathbf{k}} \hat{a}_{TE,\mathbf{k}} e^{-i\omega_\mathbf{k} t} + \mathbf{E}_{TE,\mathbf{k}}^* \hat{a}_{TE,\mathbf{k}}^\dagger e^{i\omega_\mathbf{k} t} \right)$$

$$+ \sum_{\mathbf{k}} \left( \mathbf{E}_{TM,\mathbf{k}} \hat{a}_{TM,\mathbf{k}} e^{-i\omega_\mathbf{k} t} + \mathbf{E}_{TM,\mathbf{k}}^* \hat{a}_{TM,\mathbf{k}}^\dagger e^{i\omega_\mathbf{k} t} \right).$$

$\hat{a}_{TE,\mathbf{k}}$ and $\hat{a}_{TM,\mathbf{k}}$ are the annihilation operators of the cavity TE and TM modes, respectively. $\mathbf{k} \equiv k_x \mathbf{e}_x + k_y \mathbf{e}_y = k(\cos\theta_\mathbf{k} \mathbf{e}_x + \sin\theta_\mathbf{k} \mathbf{e}_y)$ is the in-plane wave vector. The corresponding electric fields are $\mathbf{E}_{TE,\mathbf{k}} = i\sqrt{\frac{\hbar\omega_\mathbf{k}}{2\epsilon S}} \mathbf{u}_{TE,\mathbf{k}} e^{i\mathbf{k}\cdot\mathbf{R}}$ and $\mathbf{E}_{TM,\mathbf{k}} = i\sqrt{\frac{\hbar\omega_\mathbf{k}}{2\epsilon S}} \mathbf{u}_{TM,\mathbf{k}} e^{i\mathbf{k}\cdot\mathbf{R}}$, with *S* the area of the 2D cavity, $\epsilon$ the cavity dielectric constant, and $\hbar\omega_\mathbf{k}$ the photon energy.

$$\mathbf{u}_{TE,\mathbf{k}} = \sqrt{\frac{2}{L}} \sin(k_z z) \left( -\sin\theta_\mathbf{k}\, \mathbf{e}_x + \cos\theta_\mathbf{k}\, \mathbf{e}_y \right), \tag{S4}$$

$$\mathbf{u}_{TM,\mathbf{k}} = \sqrt{\frac{2}{L}} \left( \frac{k_z \sin(k_z z)}{\sqrt{k_z^2 + k^2}} \cos\theta_\mathbf{k}\, \mathbf{e}_x + \frac{k_z \sin(k_z z)}{\sqrt{k_z^2 + k^2}} \sin\theta_\mathbf{k}\, \mathbf{e}_y - i\frac{k \cos(k_z z)}{\sqrt{k_z^2 + k^2}} \mathbf{e}_z \right)$$

are the internal structures of the TE and TM modes which describe the rotation of the electric field with the direction of $\mathbf{k}$. $z \in [0, L]$ is the position in the out-of-plane direction.

The $\mathbf{k}$-dependences of $\mathbf{u}_{TE,\mathbf{k}}$ and $\mathbf{u}_{TM,\mathbf{k}}$ lead to the internal gauge structures of the cavity photon, described by a non-Abelian Berry connection which is a 2x2 matrix

$$\overleftrightarrow{\mathbf{A}}_\mathbf{k}^{(TM-TE)} = \frac{k_z}{\sqrt{k_z^2 + k^2}} \frac{\mathbf{e}_z \times \mathbf{k}}{k^2} \begin{pmatrix} 0 & -i \\ i & 0 \end{pmatrix}.$$

The two cavity photons are degenerate, and the corresponding non-Abelian Berry curvature is

$$\overleftrightarrow{\Omega}_\mathbf{k}^{(TM-TE)} = \nabla_\mathbf{k} \times \overleftrightarrow{\mathbf{A}}_\mathbf{k}^{(TM-TE)} - i\overleftrightarrow{\mathbf{A}}_\mathbf{k}^{(TM-TE)} \times \overleftrightarrow{\mathbf{A}}_\mathbf{k}^{(TM-TE)} = \frac{-k_z \mathbf{e}_z}{(k_z^2 + k^2)^{3/2}} \begin{pmatrix} 0 & -i \\ i & 0 \end{pmatrix}.$$

It's convenient to change the TM-TE to the circularly polarized basis: $\hat{a}_{+,\mathbf{k}} = \frac{\hat{a}_{TM,\mathbf{k}} - i\hat{a}_{TE,\mathbf{k}}}{\sqrt{2}} e^{-i\theta_\mathbf{k}}$, $\hat{a}_{-,\mathbf{k}} = \frac{\hat{a}_{TM,\mathbf{k}} + i\hat{a}_{TE,\mathbf{k}}}{\sqrt{2}} e^{i\theta_\mathbf{k}}$. We can then write down the internal structure of the circularly polarized basis as

$$\mathbf{u}_{+,\mathbf{k}} = \frac{e^{i\theta_\mathbf{k}}}{\sqrt{L}} \left( \left( \frac{k_z \cos\theta_\mathbf{k}}{\sqrt{k_z^2+k^2}} - i\sin\theta_\mathbf{k} \right) \sin(k_z z)\, \mathbf{e}_x + \left( \frac{k_z \sin\theta_\mathbf{k}}{\sqrt{k_z^2+k^2}} + i\cos\theta_\mathbf{k} \right) \sin(k_z z)\, \mathbf{e}_y \right.$$
$$\left. - \frac{ik}{\sqrt{k_z^2+k^2}} \cos(k_z z)\, \mathbf{e}_z \right), \tag{S5}$$

$$\mathbf{u}_{-,\mathbf{k}} = \frac{e^{-i\theta_\mathbf{k}}}{\sqrt{L}} \left( \left( \frac{k_z \cos\theta_\mathbf{k}}{\sqrt{k_z^2+k^2}} + i\sin\theta_\mathbf{k} \right) \sin(k_z z)\, \mathbf{e}_x + \left( \frac{k_z \sin\theta_\mathbf{k}}{\sqrt{k_z^2+k^2}} - i\cos\theta_\mathbf{k} \right) \sin(k_z z)\, \mathbf{e}_y \right.$$
$$\left. - \frac{ik}{\sqrt{k_z^2+k^2}} \cos(k_z z)\, \mathbf{e}_z \right).$$

The cavity photon non-Abelian Berry connection and curvature under the circular basis are diagonal:

$$\overleftrightarrow{\mathbf{A}}_\mathbf{k}^{(+-)} = \begin{pmatrix} \mathbf{A}_\mathbf{k} & 0 \\ 0 & -\mathbf{A}_\mathbf{k} \end{pmatrix}, \quad \mathbf{A}_\mathbf{k} = \frac{\mathbf{e}_z \times \mathbf{k}}{k^2} \left( \frac{k_z}{\sqrt{k_z^2 + k^2}} - 1 \right), \tag{S6}$$

$$\overleftrightarrow{\Omega}_\mathbf{k}^{(+-)} = \nabla_\mathbf{k} \times \overleftrightarrow{\mathbf{A}}_\mathbf{k}^{(+-)} - i\overleftrightarrow{\mathbf{A}}_\mathbf{k}^{(+-)} \times \overleftrightarrow{\mathbf{A}}_\mathbf{k}^{(+-)} = -\frac{k_z \mathbf{e}_z}{(k_z^2 + k^2)^{3/2}} \begin{pmatrix} 1 & 0 \\ 0 & -1 \end{pmatrix}.$$

Here we would like to point out that the above non-Abelian Berry curvature originates from the $\mathbf{k}$-dependence of the vertical (z-component) electric field for the cavity TM mode. This is because we need three variables ($E_x$, $E_y$, and $E_z$) to fully describe a photon polarization, so TM and TE don't form a complete basis. When the vertical electric field of the TM mode becomes a constant, then only two variables ($E_x$ and $E_y$) are needed to describe the photon polarization. In this case TM and TE form a complete basis and the non-Abelian Berry curvature vanishes, as the nature of the Berry curvature lies in the projection of a full Hilbert space onto a incomplete subspace. Indeed if we assume

$$\mathbf{u}_{TM,\mathbf{k}} = \sqrt{\frac{2}{L}} \left( \frac{k_z \sin(k_z z)}{\sqrt{k_z^2 + k_0^2}} \cos\theta_\mathbf{k}\, \mathbf{e}_x + \frac{k_z \sin(k_z z)}{\sqrt{k_z^2 + k_0^2}} \sin\theta_\mathbf{k}\, \mathbf{e}_y - i\frac{k_0 \cos(k_z z)}{\sqrt{k_z^2 + k_0^2}} \mathbf{e}_z \right),$$

with $k_0$ a constant independent on **k**, then the non-Abelian Berry connection is

$$\vec{A}_{\mathbf{k}}^{(TM-TE)} = \frac{k_z}{\sqrt{k_z^2 + k_0^2}} \frac{\mathbf{e}_z \times \mathbf{k}}{k^2} \begin{pmatrix} 0 & -i \\ i & 0 \end{pmatrix},$$

which is still finite. Meanwhile the corresponding non-Abelian Berry curvature becomes $\vec{\Omega}_{\mathbf{k}}^{(TM-TE)} = \vec{\Omega}_{\mathbf{k}}^{(+-)} = \mathbf{0}$ just as expected.

*(2) AB-Polariton*

When the heterobilayer is placed at a vertical position with $\cos(k_z z) = \mathbf{0}$ and $|\sin(k_z z)| = \mathbf{1}$, the cavity photon only couples to $A_\sigma$ and $B_\sigma$ interlayer excitons with in-plane transition dipoles. The polariton dispersion consists of three doubly degenerate branches as shown in maintext Fig. 3a-b. For the $n$th branch ($n = 1,2,3$), the polariton states can be written as $|n_{+,\mathbf{k}}\rangle = \varphi_{p,n,\mathbf{k}}|+_{\mathbf{k}}\rangle + \varphi_{A,n,\mathbf{k}}|A_{\uparrow,\mathbf{k}}\rangle + \varphi_{B,n,\mathbf{k}}|B_{\downarrow,\mathbf{k}}\rangle$ and $|n_{-,\mathbf{k}}\rangle = \varphi_{p,n,\mathbf{k}}|-_{\mathbf{k}}\rangle + \varphi_{A,n,\mathbf{k}}|A_{\downarrow,\mathbf{k}}\rangle + \varphi_{B,n,\mathbf{k}}|B_{\uparrow,\mathbf{k}}\rangle$, where the internal structure of $|\pm_{\mathbf{k}}\rangle$ is $\mathbf{u}_{\pm,\mathbf{k}}$ whereas $|A_{\sigma,\mathbf{k}}\rangle$ and $|B_{\sigma,\mathbf{k}}\rangle$ have a constant internal structure independent on **k**. As the coupling coefficients between $A_\sigma$, $B_\sigma$ and the cavity photons can be set as real (Eq. (6) in the maintext), here $\varphi_{p,n,\mathbf{k}}$, $\varphi_{A,n,\mathbf{k}}$ and $\varphi_{B,n,\mathbf{k}}$ are all real.

Note that $\{+_{\mathbf{k}}, A_{\uparrow,\mathbf{k}}, B_{\downarrow,\mathbf{k}}\}$ is fully decoupled with $\{-_{\mathbf{k}}, A_{\downarrow,\mathbf{k}}, B_{\uparrow,\mathbf{k}}\}$ considering $\left\langle \mathbf{u}_{+,\mathbf{k}} \left| \frac{\partial \mathbf{u}_{-,\mathbf{k}}}{\partial k_{x,y}} \right. \right\rangle = \mathbf{0}$. The 6x6 matrix form of the polariton non-Abelian Berry connection can be decomposed into two 3x3 blocks. The non-Abelian Berry connection in $\{+_{\mathbf{k}}, A_{\uparrow,\mathbf{k}}, B_{\downarrow,\mathbf{k}}\}$ subspace is

$$\vec{A}_{\mathbf{k}}^{(AB-P)} = \mathbf{A}_{\mathbf{k}} \begin{pmatrix} \varphi_{p,1,\mathbf{k}}^2 & \varphi_{p,1,\mathbf{k}}\varphi_{p,2,\mathbf{k}} & \varphi_{p,1,\mathbf{k}}\varphi_{p,3,\mathbf{k}} \\ \varphi_{p,2,\mathbf{k}}\varphi_{p,1,\mathbf{k}} & \varphi_{p,2,\mathbf{k}}^2 & \varphi_{p,2,\mathbf{k}}\varphi_{p,3,\mathbf{k}} \\ \varphi_{p,3,\mathbf{k}}\varphi_{p,1,\mathbf{k}} & \varphi_{p,3,\mathbf{k}}\varphi_{p,2,\mathbf{k}} & \varphi_{p,3,\mathbf{k}}^2 \end{pmatrix}.$$

Since the three polariton branches of $|n_{+,\mathbf{k}}\rangle$ are well separated in energy (see Fig. 3a-b in the maintext), the $n$th polariton branch then has an Abelian Berry curvature

$$\Omega_{+,n,\mathbf{k}}^{(AB-P)} = \nabla_{\mathbf{k}} \times \left( \mathbf{A}_{\mathbf{k}} \varphi_{p,n,\mathbf{k}}^2 \right) \quad \quad \quad \text{(S7)}$$

$$= \left( (\rho_{n,\mathbf{k}} - \mathbf{1}) \frac{k_z}{(k_z^2 + k^2)^{3/2}} + \frac{\mathbf{k} \cdot \nabla_{\mathbf{k}} \rho_{n,\mathbf{k}}}{k_z^2 + k^2 + k_z\sqrt{k_z^2 + k^2}} \right) \mathbf{e}_z.$$

Here $\rho_{n,\mathbf{k}} = \mathbf{1} - \varphi_{p,n,\mathbf{k}}^2$ is the fraction of the interlayer exciton. Due to the time reversal relation between $|n_{+,\mathbf{k}}\rangle$ and $|n_{-,\mathbf{k}}\rangle$, $\Omega_{-,n,\mathbf{k}}^{(AB-P)} = -\Omega_{+,n,\mathbf{k}}^{(AB-P)}$.

*(3) C-Polariton*

When the heterobilayer is placed at a vertical position with $\sin(k_z z) = \mathbf{0}$ and $|\cos(k_z z)| = \mathbf{1}$, the cavity photon only couples to the bright interlayer exciton $|C_{b,\mathbf{k}}\rangle \equiv (|C_{\uparrow,\mathbf{k}}\rangle + |C_{\downarrow,\mathbf{k}}\rangle)/\sqrt{\mathbf{2}}$ with out-of-plane transition dipoles. The $A_\sigma$, $B_\sigma$ and the dark $C_\sigma$ exciton $|C_{d,\mathbf{k}}\rangle \equiv (|C_{\uparrow,\mathbf{k}}\rangle - |C_{\downarrow,\mathbf{k}}\rangle)/\sqrt{\mathbf{2}}$ are decoupled from the cavity photon thus can be dropped. The polariton branches can be written as $|UP_{\mathbf{k}}\rangle = \varphi_{1,\mathbf{k}}|TM_{\mathbf{k}}\rangle + \varphi_{2,\mathbf{k}}|C_{b,\mathbf{k}}\rangle$, $|TE_{\mathbf{k}}\rangle$ and $|LP_{\mathbf{k}}\rangle = \varphi_{2,\mathbf{k}}|TM_{\mathbf{k}}\rangle - \varphi_{1,\mathbf{k}}|C_{b,\mathbf{k}}\rangle$, respectively. Here $\varphi_{1,\mathbf{k}}$ and $\varphi_{2,\mathbf{k}}$ are real.

$|TM/TE_\mathbf{k}\rangle$ have the internal structures $\mathbf{u}_{TM/TE,\mathbf{k}}$ whereas $|C_{b,\mathbf{k}}\rangle$ has a constant internal structure independent on $\mathbf{k}$. So the C-polariton non-Abelian Berry connection is

$$\overleftrightarrow{\mathbf{A}}_\mathbf{k}^{(C-P)} = \begin{pmatrix} 0 & i\varphi_{1,\mathbf{k}}\frac{k_z}{\sqrt{k_z^2+k^2}}\frac{\mathbf{k}\times\mathbf{e}_z}{k^2} & i\left(\varphi_{1,\mathbf{k}}\frac{\partial\varphi_{2,\mathbf{k}}}{\partial\mathbf{k}} - \varphi_{2,\mathbf{k}}\frac{\partial\varphi_{1,\mathbf{k}}}{\partial\mathbf{k}}\right) \\ -i\varphi_{1,\mathbf{k}}\frac{k_z}{\sqrt{k_z^2+k^2}}\frac{\mathbf{k}\times\mathbf{e}_z}{k^2} & 0 & -i\varphi_{2,\mathbf{k}}\frac{k_z}{\sqrt{k_z^2+k^2}}\frac{\mathbf{k}\times\mathbf{e}_z}{k^2} \\ i\left(\varphi_{2,\mathbf{k}}\frac{\partial\varphi_{1,\mathbf{k}}}{\partial\mathbf{k}} - \varphi_{1,\mathbf{k}}\frac{\partial\varphi_{2,\mathbf{k}}}{\partial\mathbf{k}}\right) & i\varphi_{2,\mathbf{k}}\frac{k_z}{\sqrt{k_z^2+k^2}}\frac{\mathbf{k}\times\mathbf{e}_z}{k^2} & 0 \end{pmatrix}. \quad (S8)$$

We note that *TE* and *LP* branches are degenerate at $\mathbf{k}=0$, and *TE* and *UP* branches becomes nearly degenerate for large $k$ values (see Fig. 4a in the maintext). So a non-Abelian Berry curvature should be considered:

$$\overleftrightarrow{\Omega}_\mathbf{k}^{(C-P)} = \nabla_\mathbf{k} \times \overleftrightarrow{\mathbf{A}}_\mathbf{k}^{(C-P)} - i\overleftrightarrow{\mathbf{A}}_\mathbf{k}^{(C-P)} \times \overleftrightarrow{\mathbf{A}}_\mathbf{k}^{(C-P)}$$

$$= \frac{k_z \mathbf{e}_z}{(k_z^2+k^2)^{3/2}}\begin{pmatrix} 0 & i\varphi_{1,\mathbf{k}} & 0 \\ -i\varphi_{1,\mathbf{k}} & 0 & -i\varphi_{2,\mathbf{k}} \\ 0 & i\varphi_{2,\mathbf{k}} & 0 \end{pmatrix}. \quad (S9)$$

Note that $\overleftrightarrow{\Omega}_\mathbf{k}^{(C-P)}$ only contains the off-diagonal terms, so the Berry curvature expectation value for each C-polariton branch is 0.

When driven by a weak external force $\mathcal{F}$, a polariton wavepacket with a momentum-space center $\mathbf{k}$ has a form $\propto \eta_{UP,\mathbf{k}}|UP_\mathbf{k}\rangle + \eta_{TE,\mathbf{k}}|TE_\mathbf{k}\rangle + \eta_{LP,\mathbf{k}}|LP_\mathbf{k}\rangle$. Following Eq. (S2), the pseudospin vector $\vec{\eta}_\mathbf{k} \equiv (\eta_{UP,\mathbf{k}}, \eta_{TE,\mathbf{k}}, \eta_{LP,\mathbf{k}})^T$ should be the eigenstate of $\overleftrightarrow{H}_\mathbf{k} - \mathcal{F}\cdot\overleftrightarrow{\mathbf{A}}_\mathbf{k}^{(C-P)}$, where $\overleftrightarrow{H}_\mathbf{k} = \mathbf{diag}(E_{UP,\mathbf{k}}, \omega_\mathbf{k}, E_{LP,\mathbf{k}})$ gives the polariton dispersion. Up to the linear order of $F$, the perturbative eigenstates closest to $|UP_\mathbf{k}\rangle$ and $|LP_\mathbf{k}\rangle$ are

$$\vec{\eta}_{UP,\mathbf{k}} = \left(1,\quad i\frac{\varphi_{1,\mathbf{k}}}{E_{UP,\mathbf{k}}-\omega_\mathbf{k}}\frac{k_z}{\sqrt{k_z^2+k^2}}\frac{(\mathbf{k}\times\mathcal{F})\cdot\mathbf{e}_z}{k^2},\quad i\frac{\left(\varphi_{2,\mathbf{k}}\frac{\partial\varphi_{1,\mathbf{k}}}{\partial\mathbf{k}}-\varphi_{1,\mathbf{k}}\frac{\partial\varphi_{2,\mathbf{k}}}{\partial\mathbf{k}}\right)\cdot\mathcal{F}}{E_{UP,\mathbf{k}}-E_{LP,\mathbf{k}}}\right)^T,$$

$$\vec{\eta}_{LP,\mathbf{k}} = \left(i\frac{\left(\varphi_{1,\mathbf{k}}\frac{\partial\varphi_{2,\mathbf{k}}}{\partial\mathbf{k}}-\varphi_{2,\mathbf{k}}\frac{\partial\varphi_{1,\mathbf{k}}}{\partial\mathbf{k}}\right)\cdot\mathcal{F}}{E_{LP,\mathbf{k}}-E_{UP,\mathbf{k}}},\quad i\frac{\varphi_{2,\mathbf{k}}}{E_{LP,\mathbf{k}}-\omega_\mathbf{k}}\frac{k_z}{\sqrt{k_z^2+k^2}}\frac{(\mathbf{k}\times\mathcal{F})\cdot\mathbf{e}_z}{k^2},\quad 1\right)^T.$$

Note that the *UP* (*LP*) polariton feels a driving force $\mathcal{F} = \varphi_{2,\mathbf{k}}^2 \mathbf{F}$ ($\mathcal{F} = \varphi_{1,\mathbf{k}}^2 \mathbf{F}$) with $\mathbf{F}$ the external force applied on the interlayer exciton. The photon polarizations of these two states are

$$\langle\hat{\sigma}_z\rangle_{UP,\mathbf{k}} = \frac{2k_z}{\sqrt{k_z^2+k^2}}\frac{(\mathbf{k}\times\mathbf{F})\cdot\mathbf{e}_z}{k^2}\frac{|\varphi_{1,\mathbf{k}}\varphi_{2,\mathbf{k}}|^2}{E_{UP,\mathbf{k}}-\omega_\mathbf{k}}, \quad (S10)$$

$$\langle\hat{\sigma}_z\rangle_{LP,\mathbf{k}} = \frac{2k_z}{\sqrt{k_z^2+k^2}}\frac{(\mathbf{k}\times\mathbf{F})\cdot\mathbf{e}_z}{k^2}\frac{|\varphi_{1,\mathbf{k}}\varphi_{2,\mathbf{k}}|^2}{E_{LP,\mathbf{k}}-\omega_\mathbf{k}},$$

respectively. Here $\hat{\sigma}_z \equiv |+_\mathbf{k}\rangle\langle+_\mathbf{k}| - |-_\mathbf{k}\rangle\langle-_\mathbf{k}| = i|TE_\mathbf{k}\rangle\langle TM_\mathbf{k}| - i|TM_\mathbf{k}\rangle\langle TE_\mathbf{k}|$ is the photon circular polarization operator.

We now look at the wavepacket velocity $\frac{d\mathbf{r}_c}{dt} \approx \vec{\eta}_\mathbf{k}^\dagger \left( \frac{\partial \overline{H}_\mathbf{k}}{\partial \mathbf{k}} + i\left[\overleftrightarrow{H}_\mathbf{k}, \overrightarrow{A}_\mathbf{k}^{(C-P)}\right] + \overrightarrow{\Omega}_\mathbf{k}^{(C-P)} \times \mathcal{F} \right)\vec{\eta}_\mathbf{k}$. Since $\overrightarrow{\Omega}_\mathbf{k}^{(C-P)}$ only contains the off-diagonal terms, $\vec{\eta}_\mathbf{k}^\dagger \overrightarrow{\Omega}_\mathbf{k}^{(C-P)} \vec{\eta}_\mathbf{k} \times \mathcal{F} \approx 0$ up to the linear order of $F$. It is also straightforward to get $\vec{\eta}_\mathbf{k}^\dagger \left[\overleftrightarrow{H}_\mathbf{k}, \overrightarrow{A}_\mathbf{k}^{(C-P)}\right] \vec{\eta}_\mathbf{k} \approx 0$. So $\frac{d\mathbf{r}_c}{dt}$ only contains the group velocity term $\vec{\eta}_\mathbf{k}^\dagger \frac{\partial \overline{H}_\mathbf{k}}{\partial \mathbf{k}} \vec{\eta}_\mathbf{k} \approx \frac{\partial E_{UP/LP,\mathbf{k}}}{\partial \mathbf{k}}$, there is no anomalous velocity for the C-polariton.

## II. Scattering induced photon polarization for C-polariton

The Rayleigh scattering of the C-polariton $UP$ and $LP$ branches can also cause different polarizations $\langle \hat{\sigma}_z \rangle$ to develop in different scattering directions. We consider an initial state $|UP_{\mathbf{k}_0}\rangle = \varphi_1 |TM_{\mathbf{k}_0}\rangle + \varphi_2 |C_{\mathbf{k}_0}\rangle = \varphi_1 \frac{e^{-i\theta_0}|+_{\mathbf{k}_0}\rangle + e^{i\theta_0}|-_{\mathbf{k}_0}\rangle}{\sqrt{2}} + \varphi_2 |C_{\mathbf{k}_0}\rangle$ with the wave vector $\mathbf{k}_0 = k(\cos\theta_0 \mathbf{e}_x + \sin\theta_0 \mathbf{e}_y)$. It is expected that the Rayleigh scattering doesn't change the polarization of the photon constituent. Since the polarization directions of TM and TE cavity photons are $\mathbf{k}$-dependent, after being scattered to $\mathbf{k} = k(\cos\theta_\mathbf{k} \mathbf{e}_x + \sin\theta_\mathbf{k} \mathbf{e}_y)$ the state becomes

$$|\psi_\mathbf{k}\rangle = \varphi_1 \frac{e^{-i\theta_0}|+_\mathbf{k}\rangle + e^{i\theta_0}|-_\mathbf{k}\rangle}{\sqrt{2}} + \varphi_2 |C_\mathbf{k}\rangle$$
$$= \varphi_1 \cos\Delta\theta |TM_\mathbf{k}\rangle - \varphi_1 \sin\Delta\theta |TE_\mathbf{k}\rangle + \varphi_2 |C_\mathbf{k}\rangle$$
$$= (\varphi_2^2 + \varphi_1^2 \cos\Delta\theta)|UP_\mathbf{k}\rangle - \varphi_1 \sin\Delta\theta |TE_\mathbf{k}\rangle - \varphi_1\varphi_2(1-\cos\Delta\theta)|LP_\mathbf{k}\rangle,$$

which is no longer a polariton eigenstate. Here $\Delta\theta \equiv \theta_\mathbf{k} - \theta_0$ is the scattering angle. The time evolution of the scattered state is

$$|\psi_\mathbf{k}(t)\rangle = (\varphi_2^2 + \varphi_1^2 \cos\Delta\theta)e^{-iE_{UP,\mathbf{k}}t}|UP_\mathbf{k}\rangle - \varphi_1 \sin\Delta\theta \, e^{-i\omega_\mathbf{k}t}|TE_\mathbf{k}\rangle$$
$$- \varphi_1\varphi_2(1-\cos\Delta\theta)e^{-iE_{LP,\mathbf{k}}t}|LP_\mathbf{k}\rangle,$$

which gives rise to a time-dependent circular polarization

$$\langle \hat{\sigma}_z(t)\rangle = 2\varphi_1^2 \sin\Delta\theta \left[(\varphi_2^2 + \varphi_1^2 \cos\Delta\theta)\sin(E_{UP,\mathbf{k}}t - \omega_\mathbf{k}t)\right.$$
$$\left. + \varphi_2^2(1-\cos\Delta\theta)\sin(\omega_\mathbf{k}t - E_{LP,\mathbf{k}}t)\right]$$
$$= 2\varphi_1^2 \varphi_2^2 \sin\Delta\theta \left[\sin(E_{UP,\mathbf{k}}t - \omega_\mathbf{k}t) + \sin(\omega_\mathbf{k}t - E_{LP,\mathbf{k}}t)\right]$$
$$+ \varphi_1^2 \sin(2\Delta\theta)\left[\varphi_1^2 \sin(E_{UP,\mathbf{k}}t - \omega_\mathbf{k}t) - \varphi_2^2 \sin(\omega_\mathbf{k}t - E_{LP,\mathbf{k}}t)\right].$$

$\langle \hat{\sigma}_z(t)\rangle$ is the superposition of two terms with $\sin\Delta\theta$ and $\sin(2\Delta\theta)$ angular patterns, respectively. The photon polarization $\langle \hat{\sigma}_z(t)\rangle$ for an initial state $|LP_{\mathbf{k}_0}\rangle = \varphi_2|TM_{\mathbf{k}_0}\rangle - \varphi_1|C_{\mathbf{k}_0}\rangle$ can be obtained similarly.

When $E_{UP,\mathbf{k}} - \omega_\mathbf{k} = \omega_\mathbf{k} - E_{LP,\mathbf{k}}$, there is $\varphi_1^2 = \varphi_2^2 = \frac{1}{2}$ and the $\sin(2\Delta\theta)$ term vanishes. In this case $\langle \hat{\sigma}_z(t)\rangle = \sin\Delta\theta \sin(E_{UP,\mathbf{k}}t - \omega_\mathbf{k}t)$. Together with a finite cavity photon lifetime, a finite and $\Delta\theta$-dependent $\langle \hat{\sigma}_z \rangle$ for the emitted photon can emerge (see Fig. 4d in the maintext).

[1] D. Xiao, M.-C. Chang, and Q. Niu, Rev. Mod. Phys. **82**, 1959 (2010).